\documentclass[10pt,onecolumn]{article}

\usepackage{amssymb}
\usepackage{algorithmic}
\usepackage{amsfonts}
\usepackage{fancyheadings}
\usepackage{fancybox}
\usepackage{multicol}
\usepackage{wrapfig}
\usepackage{color}
\definecolor{orange}{rgb}{1,0.5,0}
\usepackage{epsfig}
\usepackage{algorithm}
\usepackage{url}
\usepackage{subfig}
\usepackage[figuresright]{rotating}
\usepackage{mdwlist}
\usepackage[cmex10]{amsmath}
\usepackage{setspace}
\usepackage{tabularx}
\usepackage{tabulary}
\usepackage[margin=22mm]{geometry}

\begin{document}

\title{A Survey and Taxonomy of Urban Traffic Management:\\ Towards Vehicular Networks}

\author{
\small Hany~Kamal$^1$, Marco Picone$^2$, Michele Amoretti$^2$\\
\small 1: Rakedet, Egypt\\
\small 2: Dipartimento di Ingegneria dell'Informazione, Universit\`a degli Studi di Parma, Italy\\
\small emails: hany.k.hassan@gmail.com, {michele.amoretti, marco.picone}@unipr.it
}

\date{}

\maketitle

\doublespacing

\begin{abstract}
Urban Traffic Management (UTM) topics have been tackled since long time, mainly by civil engineers and by city planners. The introduction of new communication technologies --- such as cellular systems, satellite positioning systems and inter-vehicle communications --- has significantly changed the way researchers deal with UTM issues. In this survey, we provide a review and a classification of how UTM has been addressed in the literature. We start from the recent achievements of ``classical'' approaches to urban traffic estimation and optimization, including methods based on the analysis of data collected by fixed sensors (\textit{e.g.}, cameras and radars), as well as methods based on information provided by mobile phones, such as Floating Car Data (FCD). Afterwards, we discuss urban traffic optimization, presenting the most recent works on traffic signal control and vehicle routing control. Then, after recalling the main concepts of Vehicular Ad-Hoc Networks (VANETs), we classify the different VANET-based approaches to UTM, according to three categories (``pure'' VANETs, hybrid vehicular-sensor networks and hybrid vehicular-cellular networks), while illustrating the major research issues for each of them. 
The main objective of this survey is to provide a comprehensive view on UTM to researchers with focus on VANETs, in order to pave the way for the design and development of novel techniques for mitigating urban traffic problems, based on inter-vehicle communications. 
\end{abstract}


\section{Introduction}
Urban traffic congestion is a major problem facing the modern world, causing losses of billions of hours every year, with a negative impact on productivity and consequently on the world's economy \cite{Miller2009}. 
According to the Urban Mobility Report \cite{David2012}, in 2011 the cost of traffic congestion (in terms of wasted time and fuel) was $121$ billion dollars, considering $498$ urban areas of the United States.
Besides, traffic congestion contributes to the climate change. 
The US Department of Transportation reported that the transportation sector accounted for about $28$\% of total US Greenhouse Gas (GHG) emissions, in $2006$, making it the second largest source of GHG emissions.
In $2011$, $56$ billion pounds of additional carbon dioxide (CO$_2$) gas were released into the atmosphere because of traffic congestion, according to report \cite{David2012}.
Vehicles traveling at $60$ Km/h emit $40$\% less carbon emissions than vehicles traveling at $20$ Km/h \cite{Ezell2010}. For these reasons, both academic researchers and industry leaders agree on the importance of ITS (Intelligent Transportation Systems) to reduce traffic congestion, with the long-term goal of mitigating environmental changes.

ITS covers several mobility aspects, including Urban Traffic Management (UTM), safe driving, and easy parking, with particular attention to greening challenges. 
ITS merges and benefits from several scientific and technical fields, such as traffic theory, image processing, sensing and communication systems design. 
UTM, in particular, is no more uniquely a civil engineering problem, because of the progress in communication technologies such as cellular, satellite positioning, Vehicle-to-Infrastructure (V2I) and Vehicle-to-Vehicle (V2V) systems.
 
\begin{figure}[ht]
\centering
\includegraphics[scale=0.3]{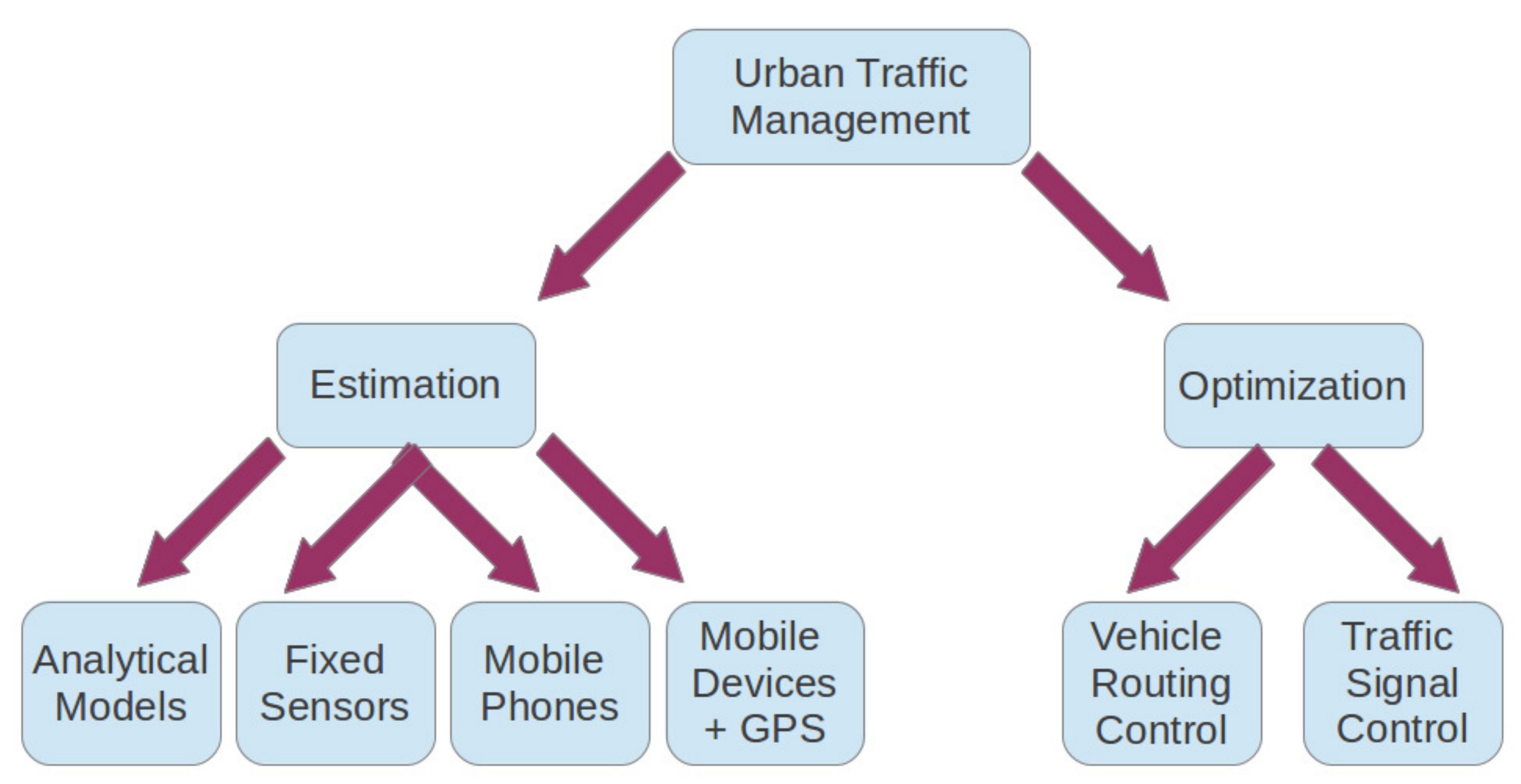}
\caption{Taxonomy of UTM categories, with classical approaches.}
\label{fig:Taxo2}
\end{figure}

UTM is usually achieved by means of a two-step process. Firstly, road traffic information is collected/estimated. Traditionally, such a collection of information is performed by means of sensors and cameras, installed along the roads. Secondly, collected information is processed (\textit{e.g.}, to obtain a view on the traffic volume, as well as routing advices, and then broadcasted by means of radio stations, electronic road panels, or public transportation web sites. The dissemination of traffic information to the drivers is expected to reduce the congestion in urban networks. 
Thus, UTM activities fall into two main categories: \textit{Urban Traffic Estimation (UTE)} and \textit{Urban Traffic Optimization (UTO)}, which have different technical approaches, represented in the bottom layer of the diagram in Fig.~\ref{fig:Taxo2}.

Traffic estimation is mainly based on the following methodologies:
\begin{enumerate}
\item analytical modeling, with Queueing Theory \cite{Woensel2006, Osorio2009} and other techniques;
\item monitoring with fixed sensors, installed under or along the roads, such as loop detectors, infrared sensors, radars or surveillance cameras, also known as Closed-Caption TeleVision (CCTV) \cite{Coifman2009};
\item monitoring with mobile phones, \textit{i.e.}, based on anonymous analysis of signaling messages and on localization of events that occur into the cellular network, such as call in progress, SMS sending, or handover \cite{Valerio2009, Calabrese2011};
\item monitoring with mobile devices combined with GPS receivers, either embedded into vehicles or into the mobile devices themselves (\textit{e.g.}, smartphones) --- a technique referred to as FCD (Floating Car Data) \cite{Messelodi2009, Tao2012, Li2009}.
\end{enumerate} 
Different approaches can be merged, in order to enhance the estimation of road traffic and to increase the accuracy of information retrieval --- we refer to these approaches as ``hybrid'' methods.

In recent years, traffic optimization has been addressed according to two main directions:
\begin{enumerate} 
\item traffic signal control, based on different techniques, \textit{e.g.}, Markov modeling \cite{Osorio2009} and Game Theory \cite{Cheng2006};
\item vehicle routing control, based on graph theory \cite{Miller2009} \cite{Scellato2010}, or assignment and insertion algorithms \cite{Fleischmann2004}.
\end{enumerate} 

In this article we review UTM techniques, with particular focus on the period $2007-2013$. 
Other survey papers appeared in the recent years, tackling similar topics. For example, Valerio \textit{et al.} \cite{valerio2009exploiting} addressed the usage of cellular networks in urban traffic estimation, while Kastrinaki \textit{et al.} \cite{kastrinaki2003survey} focused on the usage of video processing method.
The novelty of our survey is the linking of Vehicular Ad-Hoc Networks (VANETs) with UTM (in particular, UTO), compared and contrasted with ``traditional'' techniques. 
VANETs have gained the interest of both industrial and academic researchers during the last decade. 
For example, a recent book by Hartenstein \textit{et al.} provides an overview of the applications and technical aspects of VANETs \cite{HartensteinVANTE2009}. Our survey is a complementary work to such a book, with a particular focus on UTM techniques.

On the other hand, considering the recent opportunities provided by the use of the next generation of cellular access technologies in the vehicular context \cite{AranitiLTEVehicular}, our survey provides an updated view on the challenging approaches based on mobile phones, distributed and peer-to-peer overlays, VANETs and hybrid methodologies.

We would like to emphasize that our purpose is not to demonstrate that VANET-based approaches are better than traditional ones. VANETs are far from being ready for large-scale deployment. However, they are certainly interesting, as they are not only able to collect urban traffic information, but also to deal with data filtering and dissemination \cite{Killat2007}. For example, V2V and V2I communications have been proposed to obtain real-time UTE information \cite{Nadeem2004, Chen2006, Jerbi2007, Panichpapiboon2008, Garelli2011, Ma2009}. 
Moreover, UTO methods can be integrated with the VANET framework, to achieve global convergence more quickly, in a distributed manner. These approaches are discussed within a specific section of this survey.

It is worth mentioning that \emph{vehicle routing control} is completely different from \emph{routing in vehicular networks}. The former deals with routing of vehicles, while the second deals with data routing protocols in VANETs. Our survey does not address routing in vehicular networks. Other survey papers can be consulted by the reader for more details about routing protocols in VANETs \cite{lee2009survey} \cite{Chen2009}.

The remainder of the survey is organized as follows. In Sections~\ref{ute} and~\ref{uto}, we tackle the works dealt with urban traffic estimation and optimization, respectively.
In Section~\ref{vanet}, we illustrate how VANET technologies have been proposed as means for the improvement of UTM. 
In Section~\ref{model}, we discuss future research directions and challenges related to UTE and UTO.


\section{Urban Traffic Estimation} 
\label{ute}

One of the aims of traffic models is to estimate flows. These can be link flows, origin-destination flows, path flows, node flows, flows passing through a subset of nodes or links, and other. 
In addition, traffic models attempt to analyze how flows change with time and traffic intensity. Traffic flow information, estimated or collected, is used in various ITS-related applications.
This information can be integrated with traffic congestion warning systems, such as updated arrival timings of public transport, providing real-time information to the public. 
We categorize hereafter the different methods existing in the literature for urban traffic collection and/or estimation. 

Three main variables are used to model traffic streams: speed, defined as the distance covered per unit time; density, defined as the number of vehicles per unit area of the roadway; and flow, which is the number of vehicles passing a reference point per unit of time. Such variables are usually represented as a function of time, but probably the most known graph is the one which shows flow as a function of density --- which is frequently used to illustrate congestion shockwaves.

\subsection{Analytical Models}

\begin{figure}[ht!]
\centering
\includegraphics[scale=0.3]{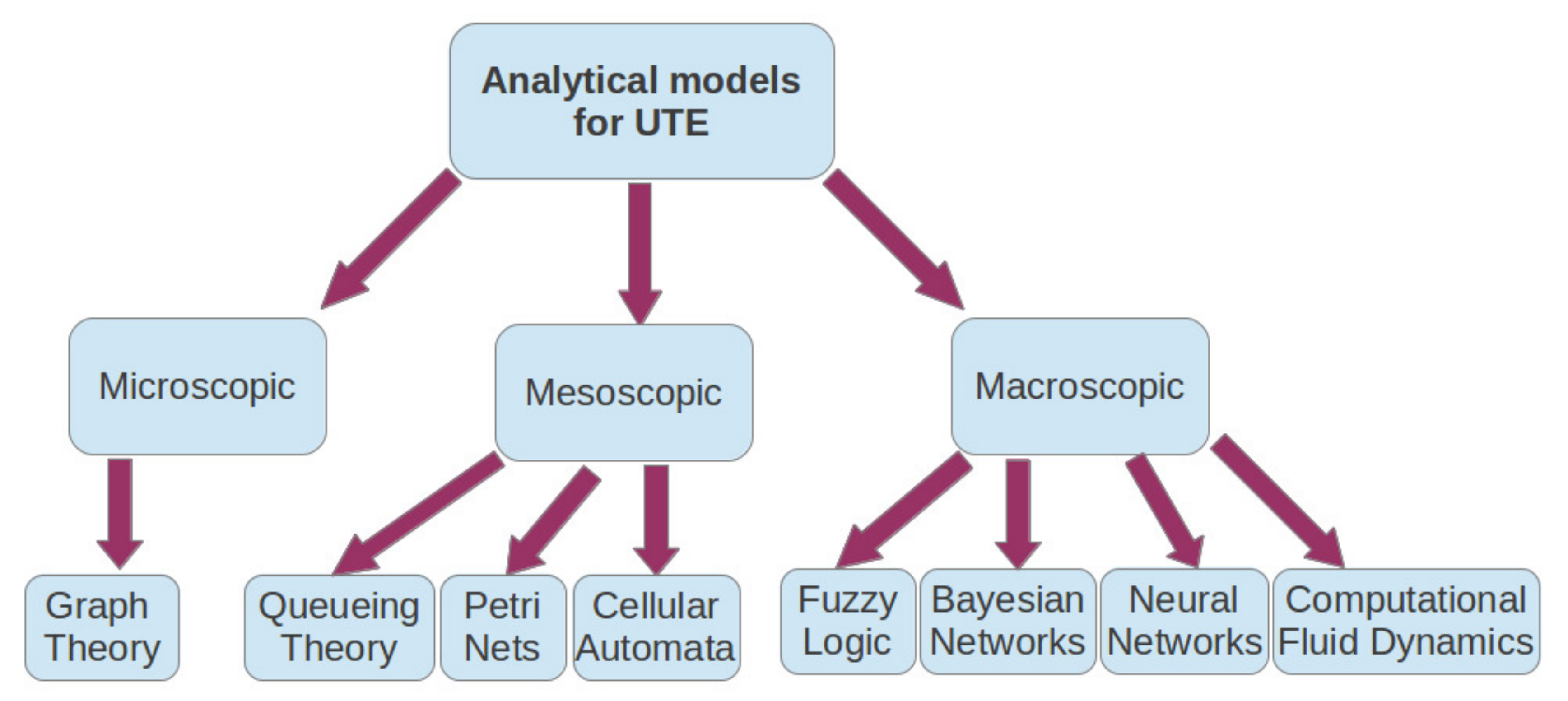}
\caption{Microscopic, mesoscopic and macroscopic analytical models for estimating the urban traffic.}
\label{fig.analytic}
\end{figure}

One of the advantages of using analytical methods is the possibility to test different scenarios --- \textit{what-if} situations --- without the need of installing devices (sensors, transceivers, etc.) in the environment. 
However, analytical modeling methods are often difficult to generalize to model large urban networks. Fig. \ref{fig.analytic} shows the different analytical modeling methods we consider in this section.

State-of-art traffic models can be classified into three groups, characterized by different granularity: 1) \textit{microscopic}, 2) \textit{macroscopic}, or 3) \textit{mesoscopic}.
Microscopic models consider single cars as entities, which move over lanes, which are parts of roads, which form a network. Car dynamics are analytically modeled, by taking into account also the effect of car density on their acceleration and speed, and the characteristics of the road. Conversely, macroscopic modeling considers traffic as a fluid, characterized in terms of density, flow, mean speed of a traffic stream, and other. Macroscopic models may integrate microscopic traffic flow models, converting the single-entity level characteristics to comparable system level characteristics.
Macroscopic models require less computation than microscopic models, and are particularly suitable for real-time applications, such as online traffic state estimation and control.
Finally, mesoscopic models fill the gap between the aggregation-based approach of macroscopic models and the individual interactions of the microscopic ones. Mesoscopic models normally describe traffic entities at a high level of detail (\textit{e.g.}, groups of cars), but their behavior and interactions are described at a lower level of detail. For example, vehicles may be grouped into packets, acting as single entities whose speed on each road (link) is derived from a speed-density function defined for that link. In another mesoscopic approach, the road is modeled as a queueing part and a running part: the vehicles traverse the running part of the road, and their speed is determined by means of a macroscopic speed-density function. At the downstream end, a queue-server is transferring the vehicles to connecting roads.

For microscopic models, several algorithms based on \textit{graph theory} have been proposed. For example, Castillo \textit{et al.} \cite{Castillo2008} illustrated two approaches --- algebraic and topological, respectively ---  to determine which subset of flows can be calculated when another subset of flows has been observed, independently of route choice probabilities.

\begin{figure}[h]
\centering
\includegraphics[scale=0.25]{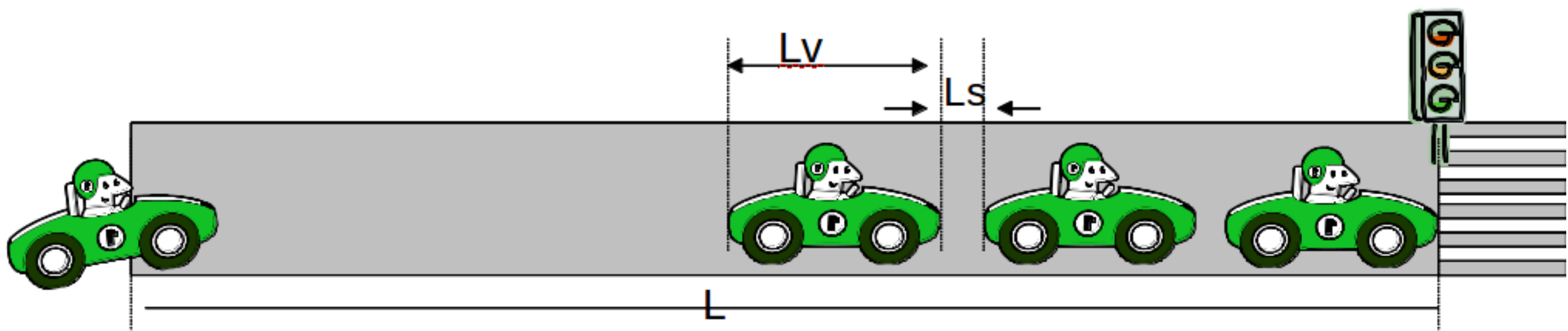}
\caption{Queue of vehicles in one lane signalized road segment.}
\label{fig:queue}
\end{figure}

With respect to mesoscopic models, \textit{Queueing Theory} has been frequently used. 
In such models, the basic idea is that the road segment offers a service to the vehicles, in a FIFO mode. The buffer size of the queue equals to the maximum number of vehicles the lane can handle.
The total time spent by a vehicle in the road segment is the sum of the service time and the waiting time. 
Fig~\ref{fig:queue} illustrates the concept, for a signalized road segment.

Woensel and Vandaele showed that, in most cases, highway traffic flows during non-congested hours are best described using a M/G/1 queueing model \cite{Woensel2006}. 
During congested hours, instead, the state dependent queueing GI/G/m models are more realistic. 
The analysis was conducted by means of a dataset collected by the ministry of transportation of the Flemish Government (Belgium). To evaluate the difference between the speeds obtained from the queueing models and the ones empirically observed, the authors adopted the Theil inequality coefficient $T$, instead of the root mean square error, because the former statistical tool does not over-emphasizes large errors. The M/G/1 queueing model gave $T = 0.03440$, regarding non-congested hours. The GI/G/3 model gave $T = 0.03516$ for morning congestion, and $T = 0.03760$ for evening congestion. All other models gave higher $T$ values.

More recently, Osorio and Bierlaire proposed an approximate queueing network model that resorts to finite capacity queueing theory to account for congested conditions in urban traffic \cite{Osorio2009}. The proposed model considers a set of intersections and analytically captures the interactions between queues. 

Also \textit{Petri Nets} are widely used to model concurrent vehicle behaviors in urban traffic models. 
In a recent work \cite{LopezNeri2010}, Lopez-Neri \textit{et al.} used the multi-level Petri net based formalism named $n$-LNS \cite{SanchezHerrera2004}, to describe the structure of urban traffic systems and their components' behaviors. 
The proposed hierarchical modeling framework allows to study the interaction of one road user with other users, static components and traffic information signals, leading to more complex emergent behaviors (such as queues, traffic jams, gridlock, green waves). The knowledge produced by the $n$-LNS model can be used to feed micro-simulators. 
The authors used an agent-based simulator and studied a network structure consisting of $38$ one-lane and one-way segments and $60$ traffic lights; each intersection had $4$ traffic lights. 
During the simulation, $1000$ vehicles were generated, and the flow-density relations were analyzed on the different segments. Four strategies for traffic light timings were compared. Unfortunately, simulation results were not validated by means of empirical data comparisons. 

Tonguz \textit{et al.} introduced a novel \textit{Cellular Automata (CA)} approach to construct urban traffic mobility models \cite{Tonguz2009}. The authors implemented a model of road intersection with traffic signal coordination, providing rules for realistic motion of turning vehicles. Due to its discrete nature, the CA model allows very fast implementation and can microscopically simulate huge networks in real-time. The authors used the New York City traffic pattern as input, and presented simulation results for: the average flow rate at an intersection, the average number of congested intersections, and the traffic density.

A Bayesian Network (BN) is a directed acyclic graph in which variables are represented by nodes and arcs between nodes represent conditional dependencies between the variables.
Most BN models used in stochastic dynamic models are based on Gaussian or mixtures of Gaussian distributions. To provide a more realistic and a positive range for the variables, Castillo \textit{et al.} have recently proposed a generalized beta-Gaussian Bayesian network (GBGN) \cite{Castillo2012}. 
The proposed model is applicable to very large networks, and was tested using three networks of increasing complexity, namely the Nguyen-Dupuis network \cite{Nguyen1984}, the Ciudad Real network with $219$ links and $590$ paths and the Vermont State network (monitored by $85$ stations\footnote{\url{http://www.aot.state.vt.us}}). 
The GBGN provides very reasonable predictions, and shows a good behavior with respect to missing data, as long as the number of affected links is small. For example, regarding the flow, the mean root square error between simulated data and real data is below $15\%$.

Tan \textit{et al.} proposed a data aggregation approach for traffic flow prediction \cite{Tan2009}. The source time series is the traffic flow volume that is collected $24$ h/day over several years. A weekly similarity time series, a daily similarity time series, and an hourly time series are generated from the source time series, using a moving average (MA) model, an exponential smoothing (EXP) model and an autoregressive MA (ARIMA) model, respectively. The proposed data aggregation strategy uses a \textit{Neural Network}, which allows to obtain a more accurate forecast than any individual model alone. Predicted flow results are compared with real data, showing a mean absolute percentage error ranging from $6\%$ to $11\%$, for increasing values of the forecast horizon (from 1 to 3 hours).

Neural networks have been used, more recently, also by Chan \textit{et al.} for short-term traffic flow forecasting \cite{Chan2012}. The proposed approach (called EXP-LM) uses exponential smoothing to preprocess traffic flow data by removing the lumpiness, before employing a variant of the Levenberg-Marquardt (LM) algorithm to train the weights of a neural network. Results indicate that, in general, test errors obtained by EXP-LM to forecast flow  are smaller than those obtained by the other tested algorithms. In general, the test error of the EXP-LM approach is always below $13 \%$.

\textit{Fuzzy Logic} is also widely used to support the modeling and prediction of urban traffic flow. For example, Dimitriou \textit{et al.} presented an adaptive hybrid fuzzy rule-based system (FRBS) approach for the modeling and short-term forecasting of traffic flow in urban arterial networks \cite{Dimitriou2008}. The approach has the advantage of suitably addressing data imprecision and uncertainty. Indeed, the mean squared relative error for the traffic flow modeling was $3.8\%$, the best one among those of the approaches considered in this survey. It also enables the incorporation of experts knowledge on local traffic conditions within the model structure.
Srinivasan \textit{et al.} used a fuzzy input fuzzy output filter (FIFO-filter) for grouping traffic patterns with similar characteristics. In addition, the authors used a multi-layer feed-forward neural network (MLFN), optimized by means of evolutionary strategies (ES), to predict the traffic in the next time step based only on the present traffic input \cite{Srinivasan2009}. The MLFN-ES model gave the following values of mean absolute percentage error, for the traffic flow: $10.02\%$ on Saturdays, $9.44\%$ on Sundays, and an average of $8.72\%$ for the entire test set, including week day and weekend data.

\subsection{Fixed Sensors}

Advanced Traffic Information Systems (TIS), but also Urban Traffic Optimization (UTO) systems (see section \ref{uto}), require real-time estimate of the current traffic state as an input. Single-loop detectors are the most common vehicle detector. Many new, out-of-pavement, detectors seek to replace loop detectors by emulating the operation of single-loop detectors. 
Coifman and Kim \cite{Coifman2009} proposed a framework based on non-conventional techniques for estimating speed at single-loop detectors, yielding estimates that approach the accuracy of a dual-loop detector's measurements. 
From these speed estimates, the authors obtained a length-based vehicle classification, which they evaluated against concurrent measurements from video and dual-loop detectors.

The traffic state cannot be directly measured everywhere, but needs to be interpolated from incomplete, noisy and local traffic data. Commonly, volumes (flows) or average vehicle speeds are measured at certain locations in the traffic network, \textit{e.g.}, by double induction loop detectors. To estimate the total traffic state from these point measurements, interpolation between the sensors is necessary. 
One of the most widely applied estimation methods is the Lighthill-Whitham and Richards (LWR) model \cite{Lightill1955} \cite{Richards1956} with an extended Kalman filter (EKF). 
A large disadvantage of the EKF is that it is too slow to perform in real-time on large networks. To overcome this problem, van Hinsbergen \textit{et al.} proposed the localized EKF (L-EKF) \cite{vanHinsbergen2012}. In the L-EKF, several local EKFs are sequentially applied, for each cell that contains measurements, instead of using an unique EKF for the entire network. 

Another interesting approach for traffic state estimation, based on cumulative road acoustics, was proposed by Tyagi et al. \textit{et al.} \cite{Tyagi2012}. The cumulative acoustic signal, acquired from a roadside-installed single microphone, comprises several noise signals — such as tire noise and engine noise. Depending on the scenario, noise signals have spectral content that are very different from each other. Hence, they can be used to distinguish between the different traffic density states that lead to them. The authors extracted the short-term spectral envelope features of the cumulative acoustic signals and modeled their class-conditional probability distributions. Three broad traffic-density states were considered: Jammed (0–10 km/h), Medium-Flow (10–40 km/h) and Free-Flow (40
km/h and above) traffic. 

To address the problem of sensor spacing optimization, with the objective of minimizing sensor and congestion costs, Leow \textit{et al.} proposed a novel sampling theorem approach \cite{Leow2008}. Traffic information, such as flow, speed, and density, were obtained from detailed vehicle trajectories, collected by Cambridge Systematics by means of the Next Generation Simulation (NGSIM) program.\footnote{\url{http://ngsim-community.org}}
Central to the proposed approach is the Shannon sampling theorem \cite{Shannon1949}. 

In the context of the 2D traffic information signal, sampling is carried out in both time and space domains. 
From the power spectral density (PSD) of traffic information, it is possible to derive the normalized mean-square error (NMSE) associated with a particular sampling rate. By converting the sampling rate into sensor spacing, it is possible to relate the NMSE to sensor spacing.

The adoption of vision-based sensing to compute the speeds and length-based classifications of tracked vehicles is becoming an increasingly popular alternative to traditional sensors for collecting traffic data.
This is mainly due to the continuously decreasing cost of cameras and processors. 
Kanhere and Birchfield proposed a taxonomy of the different calibration methods \cite{Kanhere2010}, dividing them into two categories, depending on whether only a single vanishing point (V) or two vanishing points (VV) are available. In the presented taxonomy, the authors introduced several new methods (VVH, VVL, VLH, VVD, VWD, and VHD), in addition to the three already existing methods (VVW, VWH, and VWL). 
Here W means "known width" (of a lane, or of a vehicle), H means "known height" (of the camera), L means "known length" (either the distance between pavement markings or the length of a vehicle), and D means "known distance" (from the camera to the edge of the road). 
Results show that the methods using a known length outperform the other methods (including the VVW method).
All methods have a better performance when the camera is tilted at least by a moderate amount.

\subsection{Mobile Phones}

Using mobile phones to obtain urban traffic information is motivated by their widespread diffusion, and by the fact that they avoid the installation and maintenance of fixed sensors.
Mobile phones continuously send measurement reports to the core of the cellular network. Such measurement reports are related to the radio channel, to help the network in handling resource management tasks, \textit{e.g.}, handover decisions. 
For GSM, the output of the radio measurements, associated with each ``connected'' mobile phone, travels on the A-bis interface every $480$ ms. The A-bis interface is linking the BTS (Base Station) to the BSC (Base Station Controller).

Calabrese \textit{et al.} presented a software platform which enables real-time urban monitoring through the localization of events generated by mobile phones, \textit{i.e.}, SMS, active call, handover or local area update \cite{Calabrese2011}. Such a software platform provides four main functional modules: 
\begin{itemize}
\item \textit{localization engine}, estimating the position of each active mobile phone, by extracting the signaling messages from the A-bis interface;
\item \textit{tracking filter}, estimating the trajectory and the speed of the mobile phones; 
\item \textit{mobility state estimator}, separating the moving phones from the static phones; 
\item \textit{traffic map calculator}, giving the traffic map of the monitored area in a raster map.
\end{itemize}

\begin{figure}[h]
\centering
\includegraphics [scale=0.3] {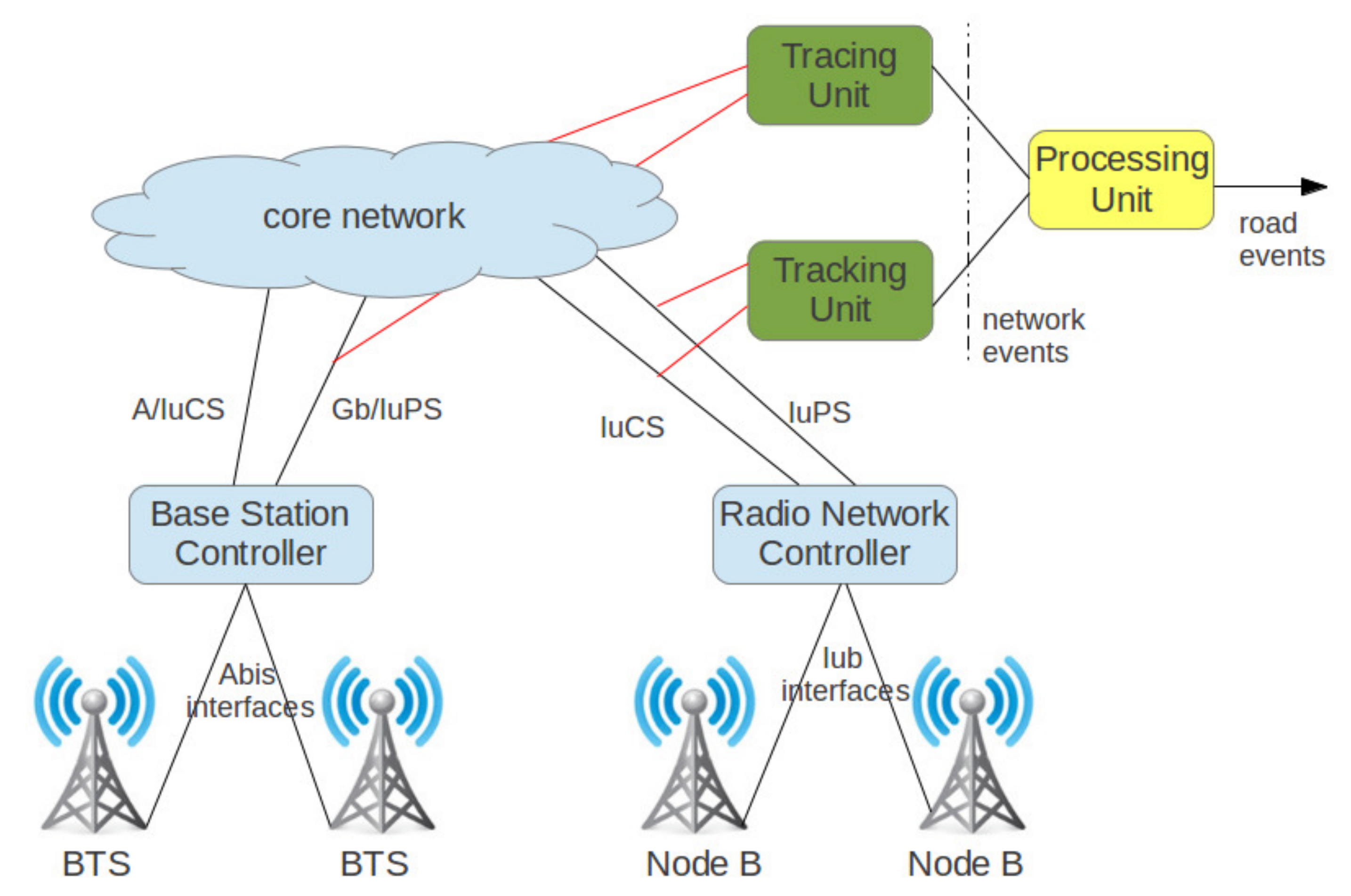}
\caption{The monitoring framework installed on a cellular operator network, based on \cite{Valerio2009}.}
\label{fig.probedata2}
\end{figure}

Valerio \textit{et al.} \cite{Valerio2009} proved experimentally that real-time road conditions map to signaling patterns in $3$G cellular network, in normal situations and in cases of car accident. 
The signaling messages are collected from Gb and IuCS interfaces, with focus on the highway scenario. 
Traces from the Iub interfaces are neglected, as high processing complexity is expected to increase in case of adding Iub traces to the traffic estimation process. 
The monitoring infrastructure used to collect the traces is composed by Tracing Unit (TU) and Processing Unit (PU). 
Network events are collected according to the 3GPP MM (Mobility Management) protocol with the use of three metrics: Location Area (LA) for terminals with Circuit Switching (CS) sessions, Routing Area (RA) for terminals with Packet Switching (PS) sessions and cell ID. Two types of events are monitored and then transmitted to the PU:
mobility-related events, such as LA update and RA update, and activity-related events, such as call setup. Fig~\ref{fig.probedata2} shows a simple architecture of the cellular system with the monitoring framework as implemented by the authors.
The news-feed of the road operator called Asfinag, in Vienna, has been monitored for potential road anomalies. On February $17$th $2009$, an accident slowed down the traffic. On the same lane of the accident, after few kilometers, there is a border between two different LAs. The analysis of the signaling traces depicts the number of users moving from the LA where the accident occurred to the following LA. The number of mobile users updating their location decreased suddenly, because they were blocked or slowed down by the accident. When the road traffic was restored, a large number of users changed LA in the same Time Interval Counter (TIC). 

It is worth mentioning that using the accumulated LA signal technique for accident monitoring in real-time has a critical time response issue. In other words, the time needed by the cellular operator to detect the accident after it has already occurred is a critical metric.
It is also to be noted that the estimation of automobile traffic using mobile phones requires a filtering process to distinguish between mobile terminals inside moving vehicles, and terminals carried by pedestrian. The filtering process is usually based on the calculated speed of connected users. This process might add complexity and overhead to the traffic estimation.

\subsection{Mobile devices and GPS receivers}
\label{sec.fcd}

The expression \textit{floating car data} refers to the data being ``continuously'' collected by a fleet of vehicles, called \textit{probes} \cite{Messelodi2009}, which are equipped with GPS receivers and GSM/GPRS transmitters to collect and send TT data to a dedicated server, in a Traffic Management Center.
 
Two types of floating-data systems are used: 1) taxi/bus FCD, where a GPS chip transmits it position via a cellular connection to the data-center, and 2) mobile-phones FCD, where private mobile phones equipped with GPS (located inside the moving vehicles) participate into providing floating-data \cite{Tao2012}. The first type faces the limitation of ``quasi-fixed'' moving patterns of taxis and buses. The second type needs to deal with particularities regarding privacy issues.
Unlike the approach based on mobile phones, FCD (or probe data) gives few though precise information samples about TT in specific road segments. 
The samples collected from probe cars are then used to obtain a global vision of the traffic in the network with the help of analytical methods.

Li \textit{et al.} used historical probe car data to discover congestion points \cite{Li2009}. The congestion level is evaluated based on real maximum TT between links. Data were collected in Beijing by $7000$ taxis, over 57 days. The study followed four steps: 
\begin{enumerate}
\item \textit{building} a database of real travel speed of links: using $7000$ taxis, equipped with GPS receivers;
\item \textit{classifying} the data according to working days and week-end, working-hours and non-working hours;
\item \textit{identifying} congestion levels with respect to the previous day and time classifications (congestion levels are: smooth, light congested and congested);
\item \textit{identifying} congestion links with an evaluation model which compares the average speed of a link with the average speed of nearby links.
\end{enumerate}

More recently Kong \textit{et al.} proposed and compared two methods for urban traffic estimation based on probe vehicles: 1) a curve-fitting-based method and 2) a vehicle-tracking-based method \cite{Kong2013}. 
The average speed of the vehicles --- which run on a specific road section during a period of time --- is used as a metric to represent the traffic state.
The authors compare the two methods regarding the estimation accuracy and the operation speed.
Results show that the tracking-based method achieves higher estimation accuracy, but slower operation speed compared with the curve-fitting technique. Using the tracking-based method, more than $80$\% accuracy can be achieved, whereas the accuracy of the curve-fitting method is always between $70$\% and $80$\%. For both methods, the accuracy is higher during off-peak hours than during peak hours.

As the distribution of probe vehicles is uneven over space and time, this causes the gathered data to be ``incomplete''. To solve the issue of missing data from probe vehicles, Zhu \textit{et al.} proposed a compressive sensing based algorithm \cite{Zhu2013}, which achieves a minimum-error estimation of the traffic condition matrix (whose rows represent time slots and columns represent road segments). The considered metrics is the average speed of traffic flow on a given link at a certain time slot. The compressive sensing theory exploits the hidden structures, based on the fact that real world data often contain redundancy. A measurement matrix, extracted from the measurements performed by the probe vehicles, is used as an input to the algorithms. Presented results show that the estimate error is as low as $20$\% even for a missing probe data of $80$\%.

During the last decade, several ICT researchers designed and analyzed peer-to-peer (P2P) \cite{Amoretti2009} algorithms and overlays for different purposes, such as file sharing, social application, live and on demand streaming and decentralized geolocation services. Distributed localization is a clear example of geocollaboration service, where active users can efficiently discover and disseminate information about existing services (such as a video stream from a webcam, storage volunteers, and other), geo-referenced objects and information (\textit{e.g.}, gasoline station, accident, bad surface conditions, traffic jams). 
These approaches, combined with mobile phones, allow to build distributed applications to efficiently harvest measuring reports, traffic information and TT in the regions of interest. They are usually implemented by recursively dividing the 2D space into smaller areas in order to assign responsibilities for region of space to peers. Instead of employing a number of centralized servers (either dedicated  or  selected among participating nodes) to carry the load for the entire network, every node shares the load of indexing and searching data that refers to its area.  

GeoP2P is an architecture that performs a hierarchical partitioning of the 2D geographic space, adopts a fully decentralized peer-to-peer overlay scheme, with overlay maintenance and query routing \cite{geop2p}. The system consists of large number of peers, distributed across a 2-dimensional space with rectangular boundary.  Each peer is responsible for providing information which is relevant to its location. A peer can be associated to a single sensor, such as a surveillance camera, or to a database which contains information about the local environment, such are a hotel, or a gas station.
Any peer can be interested in any region in the space and send a query. The purpose of the overlay network is to route queries to relevant peers.

Rybicki \textit{et al.}, with Peers on Wheels \cite{Rybicki2007}, and more recently with PeerTIS \cite{Jedrzej2009}, proposed P2P architectures where participating cars are peers organized in a Distributed Hash Table (DHT), to receive and distribute useful information to improve the vehicle TT using dynamic route guidance. In PeerTIS, roads are divided into segments, each with a unique ID that is used as key in the DHT. The main idea is that each node is responsible for a certain part of the ID space and, consequently, for a certain number of road segments. Up to now, one of the troubling issues is the fact that obtaining full information about planned and alternative routes is expensive in terms of bandwidth consumption. In PeerTIS this issue is addressed by means of a modified DHT. 

A different P2P approach called Distributed Geographic Table (DGT) was proposed by Picone \textit{et al.} \cite{Picone2011a, Picone2011b}. The DGT is a structured overlay scheme where each participant can efficiently retrieve node or resource information (data or services) located near any chosen geographic position. In this approach, the responsibility for maintaining information about the position of active peers is distributed among nodes. Consequently, a change in the set of participants causes a minimal amount of disruption. A very important aspect of this approach --- and, generally, of real-time localization systems --- concerns security and privacy. In the DGT, the only data that are shared among available peers are: a unique ID, IP address, port for the communication (common for any P2P approach) and the GPS coordinates of the node. 
The capability of finding peers that are active and close to a specific geographic position is obtained without adding any kind of personal data. Sensitive or potentially critical information may be added by applications based on DGT, however this is an issue related to the application layer.
Picone \textit{et al.} illustrated the first prototype of a DGT-smartphone-based Traffic Information System, called D4V, allowing each participant vehicle to efficiently discover traffic related information or services located near a geographic location of interest \cite{Picone2012D4V}, with high coverage percentage and low bandwidth usage.

\subsection{Hybrid methods}

Blandin \textit{et al.} \cite{Blandin2009} proposed arterial TT estimation over a road segment using FCD, machine learning techniques and convex optimization. Using the known TTs of subset of vehicles (probe cars), the TTs of all vehicles is estimated. A kernel regression method was used to obtain non-linear estimate of travel time (TT) on arterial road segment. Convex optimization was used to enhance the non-linear estimation obtained via Kernel regression. Based on the knowledge of subset of pairs, \textit{i.e.}, entry time at the road segment and TT on the same road segment, an estimation of the TT is performed for all entry time values.
 
Herring \textit{et al.} \cite{Herring2010} modeled the evolution of arterial-traffic as a Coupled Hidden Markov Model (CHMM), with the objective of estimating TT probability distributions (of each link) as well as predicting the short-term evolution of travel times. In the proposed model, the state of the Markov chain represents a discrete traffic state, \textit{i.e.} congested/saturated. Thus, the authors represented the spatio-temporal evolution of traffic in different links and at different time intervals. An algorithm to estimate traffic with FCD obtained measurements was also presented. Data measurements from probe vehicles were collected by $500$ taxis in San Francisco from random locations at random times, with probe vehicles sending GPS data to a dedicated server every minute. An expectation maximization algorithm has been used to learn the parameters of the CHMM. Given the state of the links of the network over a period of time, the parameters of the model (for example the state transition matrix) 
were estimated.

\subsection{Discussion}

Table \ref{sumEstim} summarizes the basic techniques used for urban traffic estimation, emphasizing the advantages and limitations of each method. Analytical methods are the less expensive to implement, but they produce sub-optimal results and have clear limitations, when dealing with large networks. To simulate large road networks in reasonable times, mesoscopic and macroscopic models are viable solutions. Fixed sensors are costly and sensitive to noise. Thus, their usage is limited for large roads. Mobile phones have the advantage of being highly popular and the cellular networks have a wide coverage, allowing to build cost-effective UTE services.
However, distinguishing among terminals held by pedestrians and terminals located inside moving vehicles requires an additional filtering process. More powerful mobile devices, such as smartphones and tablets, in some regions are getting even more popular than cellular phones. They are provided with many sensors, as well as GPS receivers. For these reasons, they will probably become the most used UTE support platform, 
maybe integrated with analytical models. Indeed, hybrid solutions have the challenge of optimally distributing the computational load among mobile devices and centralized servers. 
Further discussion is postponed to section \ref{model}.

\begin{table*}[ht]
\centering
\scriptsize
\caption{Summary and comparison among the different approaches to UTE.}
\label{sumEstim}
\begin{tabulary}{0.001\textwidth}{ | l | c| c| c| }

  \hline      
  \hline      
  \textbf{Methods} & \textbf{Advantages} & \textbf{Disadvantages}  & \textbf{Articles}  \\ 
  \hline      \hline   
  
  Analytical models 	& They allow to test different scenarios, & They are often topology-specific &\cite{Osorio2009,Castillo2008}\\
	                		& what-if situations, without installing  & and limited to small networks. &\cite{LopezNeri2010, Castillo2012, Tonguz2009},\\
			                & sensors in the environment. 		        & They cannot provide exact and optimal solutions.  & \cite{Tan2009,Dimitriou2008,Srinivasan2009,Chan2012}\\
			
   \hline 
  Fixed sensors  & Through the collection of sampled real-data, & They have high installation and maintenance cost.     &\cite{Coifman2009}   \\
			           & real-data traffic estimation can be          & They are highly sensible to noise  &\cite{Lightill1955,Richards1956,vanHinsbergen2012}, \\
			           & precise (depending on the sampling rate).    & and have limited coverage. They are suitable 	&\cite{Tyagi2012,Leow2008,Shannon1949,Kanhere2010} \\
			           & 																							& only for highways and main urban roads.  &\\
   \hline 
  Mobile phones 	& This approach benefits from the high popularity & It necessitates a filtering process  &\cite{Valerio2009,Calabrese2011}  \\
			& of mobile phones and avoids            	& to distinguish between terminals held by & 	\\
			& the installation of road-sensors.		& pedestrians \& terminals in vehicles. &		\\
			& 		   					& It is less accurate than GPS.		 &			\\
   \hline 
  Mobile devices & This solution provides real-time and accurate   & It has a limited coverage (due to the & \cite{Messelodi2009, Tao2012, Li2009}   \\
  and			       &  information related to specific road-segments. & limited number of probe car samples)	 & \cite{Kong2013,Zhu2013,geop2p,Rybicki2007} \\
  GPS receivers	 & 						                                   & as well as privacy related constrains	& \cite{Jedrzej2009,Picone2011a,Picone2011b,Picone2012D4V}\\

   \hline 
  Hybrid methods  	& These approaches benefit from integrating   	& Proper design is needed to balance  	& \cite{Blandin2009, Herring2010}   \\
  			& different methods, allowing to obtain  	& computational loads, for getting the maximum 	& \\
  			& near optimal solutions while reducing costs. & benefit from the different methods.  	& \\

   \hline 
   \hline 
\end{tabulary} 
\end{table*}


\section{Urban Traffic Optimization} 
\label{uto}

In this section we explore the different techniques used to address Urban Traffic Optimization (UTO), either with traffic light control (fixed-time method versus adaptive method), or with routing algorithms for vehicles.
Other approaches, such as the static reconfiguration of one-way traffic \cite{SalcedoSanz2013}, are not covered here, as they are considered to be less effective than traffic light control and vehicle routing.

In the USA, according to the National Electrical Manufacturers Association (NEMA), a different number must be assigned to each allowed \textit{movement}, in a signalized intersection.
Movements are often lumped together to run at the same time. These sets of movements are known as \textit{phases}. There may be more than one movement served in a phase, but at least one NEMA movement number must be assigned. It is common to set the phase number equal to the lowest through movement number in the phase.

\begin{figure}[h]
\centering
\includegraphics[scale=0.3]{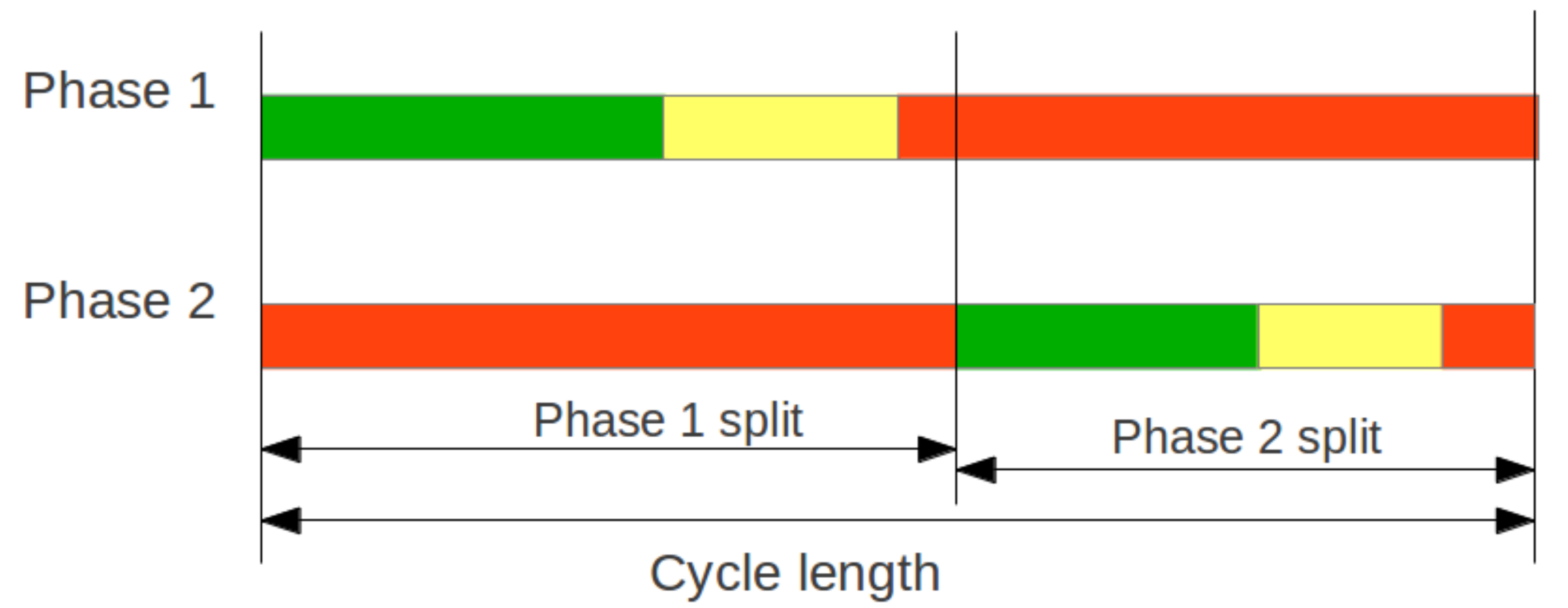}
\caption{Structure of a two-phases traffic light.}
\label{light.phase}
\end{figure}

Traffic light optimization is mostly about making a joint decision on the duration of signal phases associated to each signalized intersection in the network. Two methods are adopted: \textit{fixed-time} and \textit{adaptive-time}.
With the fixed-time method, the cycle time and the duration of one of the light (either red or green) is considered to be fixed. 
The other light's duration is derived according to the assumptions of the study. Once the optimal durations are obtained, they remain fixed --- no matter the changes of the urban-traffic.
In the adaptive method, durations are changed according to traffic conditions. In other words, the fixed-time method depends on the history of traffic statistics, while the adaptive method depends on the real-time traffic information.
Fig.~\ref{light.phase} shows the structure of a two-phases traffic light.

On the other hand, vehicle routing algorithms are techniques mainly developed to support fleet management operations, such as courier services, pickup-delivery services or taxis. The main purpose of such techniques is to provide the optimal route to drivers, based on shortest-path calculations and other constrains such as time intervals. 
Sometimes, shortest-path calculations are combined with traffic forecasts (\textit{e.g.}, TT estimations). Vehicle routing algorithms can be used also for evacuation operations in emergency situations --- in particular, to assign the optimal flow to each street, with the objective of minimizing the evacuation time, either in a centralized manner \cite{HamzaLup2005}, or in a distributed manner \cite{Goldstein2008}.

\subsection{Traffic signal control}

Traffic signal control systems are either centralized or decentralized. The former approach requires a global model of the road network, with simulated or real traffic flows. It is difficult to develop a mathematical model of an intersection for calculating the green time required for the specific traffic demand, because of the non-stationary characteristics of vehicular flow at intersections. Furthermore, the presence of signals in neighboring intersections causes platoon formation (pseudorandom behavior), limiting the usage of stochastic control models. 
Thus, the alternative approach is to deploy a distributed control architecture where each intersection is individually controlled by an intelligent agent, which autonomously decides the desired signal control policy, based on local and communicated congestion information from other agents at intersections connected to the outgoing links.

The Traffic-responsive Urban Control (TUC) approach \cite{Diakaki2002}, which has been successfully implemented in several large networks in Europe and North America, is still an important reference. TUC incorporates a predetermined plan of fixed greens for each stage at each signalized junction. Extensive investigations have shown that TUC's sensitivity to the particular adopted fixed plan is minor under high-demand conditions. In contrast, when demands and queueing are low, TUC's split decisions are close to the utilized fixed plan. 

Kouvelas \textit{et al.} \cite{Kouvelas2011} proposed a real-time version of the traditional rules by Webster and Cobbe \cite{Webster1966}, which have been extensively used by traffic engineers in the last
$50$ years, for the design of fixed-time splits under known (historical) constant demands. The derived real-time method is efficient, as long as traffic conditions are undersaturated, but it does fail when queues start to form in network links. Therefore, a hybrid approach was proposed, whereby signalized junctions are controlled by the real-time Webster-type demand-driven strategy as long as traffic conditions are under-saturated. A switch to TUC is made when traffic conditions are close to saturation. Under certain conditions (\textit{e.g.}, low but time-varying demands), the proposed extended strategy improves over the original TUC version.

Fang and Elefteriadou \cite{Fang2008} proposed a real-time signal control technique based on \textit{Dynamic Programming (DP)}. The signal plan is formulated as a decision network, where every $2.5$s a signal phase (using NEMA phase numbering) is switched or remains unchanged.
The DP algorithm optimizes the signal plan, based on the vehicle information obtained from the loop detectors that are set at certain distance from the stop line. 
The approach was applied to a simulated signalized diamond interchange instrumented with loop detectors. A comparison study showed that the adaptive signal provided by the DP algorithm can handle the demand fluctuations more effectively than two other offline optimization packages, namely, PASSER III (by Texas Instruments) and TRANSYT-7F (by McTrans).

To determine the fastest paths in freeways, Miller proposed different algorithms based on \textit{Graph Theory} \cite{Miller2009}. 
The highway network model is a directed weighted graph, where the nodes represent the intersections and the edges represent the sections of highway connecting two nodes. The weight of an edge represents the amount of time needed to travel from one node to the next node. 
The proposed algorithms focus on efficiency, in terms of reduced latency. 
It is worth mentioning that the work done by Miller focuses on highway networks, which is different from a signalized-network; \textit{e.g.}, there is much less number of intersections. Besides, the author assumes that the fastest-path is equivalent to the shortest-path, while paying less attention to traffic and waiting time in queues.

Soares and Vrancken \cite{Soares2012} observed that the dynamic behavior of a group of traffic signals controlling a network of intersections is a complete discrete event system which can be modeled by \textit{Petri Nets} (using a bottom-up strategy).  
Detectors send information about the traffic state to the controller, which sets the traffic signals to regulate the traffic. 
Model verification was based on several simulations using a token player algorithm. Afterwards, structural analysis was performed by applying the invariant theory to the Petri Net model.
Using Reachability analysis it is possible to find whether an unsafe state that could cause an accident can be reached.

Lin \textit{et al.} \cite{Lin2011b} introduced a macroscopic urban traffic model and a set of Mixed-Integer Linear Programming (MILP) controllers. The adopted methodology is \textit{Model Predictive Control (MPC)}, which repeatedly solves optimization problems online, finally deriving a sequence of control decisions. 
The designed MILP controllers were evaluated by means of simulations, with a grid network including four intersections. Although the MILP algorithms resulted in being very time efficient, for the studied case, in general, MILP problems are NP-hard. 
For larger urban traffic networks, and when the size of the MILP problem becomes too large, other control structures should be adopted. 
Such a challenging problem was studied in a subsequent article by the same authors \cite{Lin2012}. When the number of controlled intersections gets larger, the optimization problems of model-based control strategies (including MPC) become too computationally complex to be solved online. To improve the real-time feasibility, the following methods can be considered. First, dividing the network into small subnetworks and building distributed controllers. Second, solving the optimization problem offline, such as optimizing a feedback regulator offline and using it with real-time measured traffic states to derive control decisions. Third, finding efficient solution methods for the online optimization problems. Fourth, reducing the computational complexity of the control model for urban traffic networks. Lin \textit{et al.} focused on the latter aspect, in which they simplified the nonlinear traffic prediction model to reduce the online computation time. 

Zegeye \textit{et al.} \cite{Zegeye2012} proposed a Receding-Horizon Parametrized (RHP) macroscopic approach, which combines the advantages of conventional MPC (\textit{i.e.}, prediction, adaptation and handling constraints, multi-objective criteria, and nonlinear models) with those of state feedback controllers (\textit{i.e.}, faster computation speed and easier implementation). 
The proposed control approach was studied using a simulation-based case study on a part of the Dutch A12 freeway.
Results demonstrate that the proposed RHP traffic controller performs almost the same as conventional MPC.

Gokulan and Srinivasan \cite{Gokulan2010} proposed the Geometric Fuzzy Multi-Agent System (GFMAS), a distributed architecture of agents.
Each agent, which is associated to a specific intersection, receives data from local sensors, and directly calculates the desired green time for a phase, based on the averaged flow rate, queue count collected by the data-collection module and neighboring intersection status collected by the communication module. 
The architecture was evaluated in simulation, using a section of the central business district in Singapore, with signal phasing sequence, signal plans, and traffic count data obtained from the Land Transport Authority of Singapore. A comparison with benchmark signal controls GLIDE and HMS showed that GFMAS signal control outperformed them under all the simulation scenarios and was capable of alleviating the congestion experienced at the intersections.

Another decentralized architecture was proposed by De Oliveira and Camponogara \cite{deOliveira2010}. Their framework for multi-agent control of linear dynamic systems decomposes a centralized model predictive control problem into a network of coupled and small sub-problems that are solved by the distributed agents. 
Each agent senses and controls the variables of its sub-system, while communicating with agents in the vicinity to obtain neighborhood variables and coordinate their actions. A well-crafted problem decomposition and coordination protocol ensures convergence of the agents' iterations to a global optimum of the MPC problem.
Simulation analysis showed that multi-agent MPC can achieve comparable performance to the TUC approach \cite{Diakaki2002}, in representative scenarios implemented with the Aimsun simulator.

\subsection{Vehicle Routing Control}

The Vehicle Routing Problem (VRP) can be defined from two different point of views: 
\begin{enumerate}
 \item as seen by an individual driver who desires to reach a certain destination $t$ starting from a source point $s$;
 \item as seen by a ``decision maker'' who manages a fleet of vehicles providing a service, such as delivery of goods or patient transportation to hospitals.
\end{enumerate}

For the first case, the most widely adopted solution is based on a GPS receiver with a route-selection algorithm, to localize the driver and inform her/him about the ``best'' route. 
The road network is categorized as a \textit{static road network}. The problem of finding the ``best'' route is formulated using \textit{Graph Theory}, where the objective is to find the shortest path between the source $s$ and the destination node $t$.
The algorithm should not have excessive memory requirements and should be fast enough to permit a practical use to the drivers. 
The most classical and simple methods are the Dijkstra's algorithm and the A* algorithm for geometric directed search. 
Other algorithms, such as the greedy best-first-search algorithm, are faster than Dijkstra search. However, these types of algorithms, based on heuristics, do not guarantee to find the shortest path.
Moreover, partitioning methods can be used to divide large graphs into smaller ones, hence enhancing the performance through the acceleration of the computation time \cite{Mohring2005}.

In line with this first formulation of the problem, Scellato \textit{et al.} proposed a routing method based on local decisions at the vehicle level \cite{Scellato2010}, assuming that only local knowledge about congestion is available and that drivers know the shortest paths to their destinations. The network is modeled using a weighted graph where the weight of each link (edge) is a function of the link's length.
An algorithm based on Cellular Automata is used --- more precisely, a Nagel-Schreckenberg model. 
At each node (intersection), a minimization problem is to be solved by the vehicle, while considering the nearby congestion, represented by a factor, and the shortest path, in order to choose the next node as a hop on the path to destination $t$. 

\begin{figure}[h]
\centering
\includegraphics [scale=0.3] {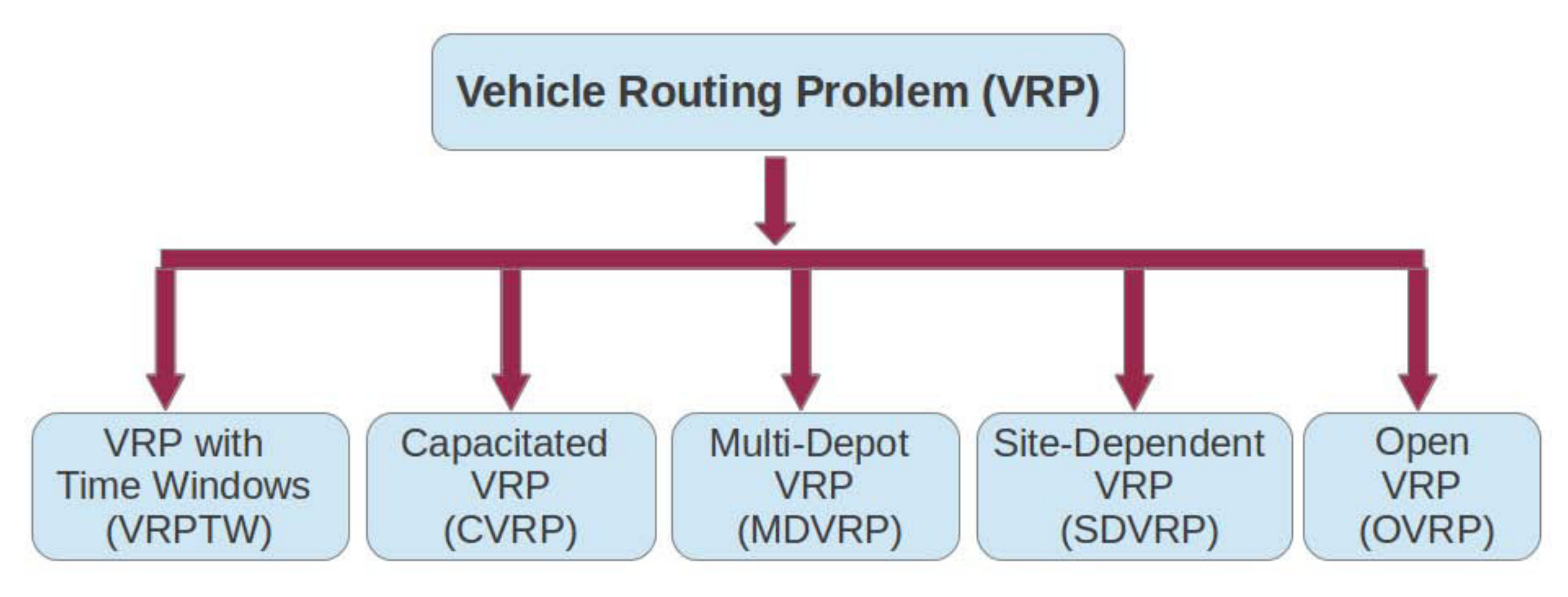}
\caption{Five categories of the Vehicle Routing Problem (VRP), as seen from a decision-maker for fleet-management operation.} 
\label{fig:vrp}
\end{figure}

For the second formulation, the problem is much more complicated. Several vehicles are to be directed to different destinations, according to the customer needs. 
The ``decision maker'' needs to find the most convenient route for each vehicle. 
The high dynamism of the problem, as well as the presence of delivery deadlines, are the most challenging issues.  
The customer's orders might arrive at any random instant. The arrival of a new order will affect all previously ``optimized'' assignment of vehicles. 
There are five main sub-categories of this VRP formulation, as seen from a decision-maker for fleet-management operation, listed below and illustrated in figure~\ref{fig:vrp}.
\begin{enumerate}
 \item Capacitated Vehicle Routing Problem (CVRP) -- Goods have to be delivered to a set of customers with minimum-cost of routes. The vehicles are supposed to start and end their trip at a certain depot. 
	The vehicles are homogeneous and having a certain capacity. 
 \item Vehicle Routing Problem with Time Windows (VRPTW) -- An extension to the CVRP by associating time windows. The customer must be served during the interval defined by the time window.
 \item Multi-Depot Vehicle Routing Problem (MDVRP) -- Extends the CVRP by allowing multiple depots.
 \item Site-Dependent Vehicle Routing Problem (SDVRP) -- Vehicles do not need to have the same capacity.
 \item Open Vehicle Routing Problem (OVRP) -- Vehicles do not need to return to the depot, so a route ends as soon as the last customer is served. 
\end{enumerate}

Pisinger \textit{et al.} proposed a solution to the previous mentioned categories using an Adaptive Large Neighborhood Search (ALNS) heuristic method \cite{Pisinger2007}.
The ALNS method, originally suggested by Gehring and Homberger \cite{Gehring1999}, uses less vehicles compared to other competing heuristics, for the large-scale VRPTW.
In each iteration of the ALNS main loop, a couple of \textit{destroyed} and \textit{repaired} neighbors is randomly created.
A destroyed neighbor is created by removing a number of orders while assigning them to the \textit{request bank}. The \textit{request bank} contains non-served orders. Different types of destroy can be used, such as the \textit{critical destroy}, where a customer is removed such that the cost of the resulting solution is minimal.
The repaired neighbor is created by inserting a number of orders from the \textit{request bank} to the routes. A new solution is accepted based on simulated annealing method, where a standard exponential cooling rate is used.
The objective is to find a feasible set of routes for the vehicles, so that all orders are served, and the overall travel distance is minimized.

The different categories listed above can be transformed into a \textit{pickup-delivery problem with time windows (PDPTW)}, where a number of customer orders are to be served by a given set of vehicles \cite{Pisinger2007}. Such a problem, in most cases, is represented by a graph, where the pickup and delivery nodes are the vertices and the routes are the edges. A length and a travel time are associated with each edge. For each vehicle, a directed sub-graph is formed. The travel cost of a given route is a function of the distance traversed by the vehicle. Solving the VRPTW problem has the objective of minimizing the total cost (over all routes) subject to the following limitations: time frame, capacity of vehicles, and number of vehicles.

\begin{figure}[h]
\centering
\includegraphics [scale=0.25] {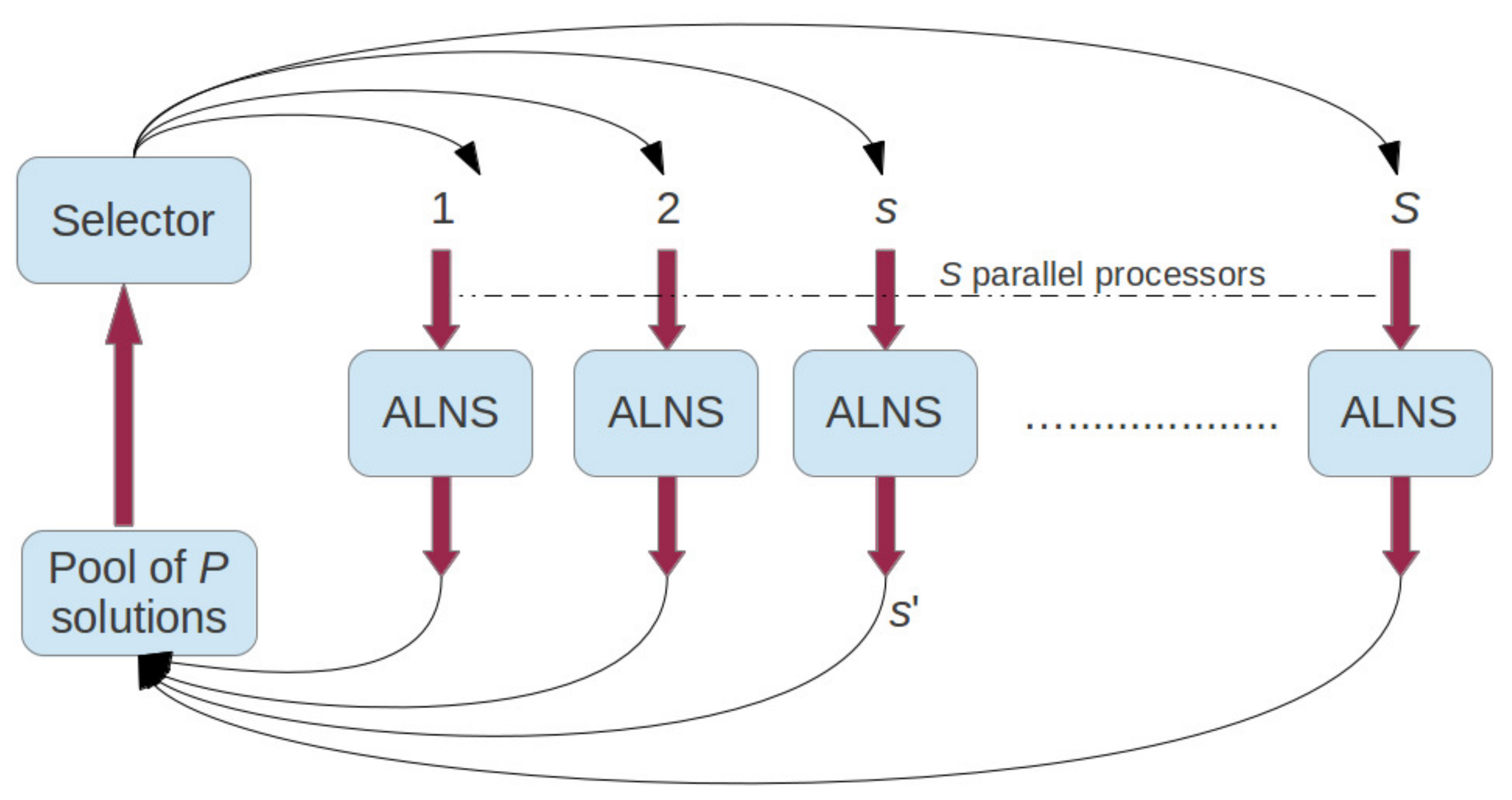}
\caption{The parallel Adaptive Large Neighborhood Search (pALNS) concept.} 
\label{fig.pALNS}
\end{figure}

Pillac \textit{et al.} addressed the VRP  for fleet of vehicles \cite{Pillac2012} by focusing on dynamic and deterministic routing --- specifically, on the Dynamic Vehicle Routing Problem with Time Windows (D-VRPTW). 
The service is to be accomplished for a set of customers over a single day period, where each customer must be served within a given time frame. While a set of customers is known beforehand, new customers may appear during the day. 
A parallel Adaptive Large Neighborhood Search (pALNS) was proposed to tackle the D-VRPTW. The pALNS algorithm allows to find an initial solution that is then re-optimized whenever a new customer request arrives.
The parallelization scheme spreads the computations across independent processors.
At each master iteration of the pALNS algorithm, a subset of $S$ solutions is randomly selected and distributed over $S$ independent processors.
Each processor executes the ALNS algorithm as described above (using destroy and repair operators). The output of each processor is added to a pool of ``optimized`` solutions.
The pool has a predefined size $P$, where $P\geq S$. Fig~\ref{fig.pALNS} provides an illustration of the pALNS concept.

A filtering procedure (represented by the \textit{selector} block in Fig~\ref{fig.pALNS}) ensures that the pool contains at most $P$ solutions. A \textit{retain} method is deployed, to always have the \textit{best} solutions in the pool, using a fitness function. 

Kanamori \textit{et al.} proposed a traffic congestion management technique based on \textit{stigmergy} and on inter-vehicle communications (see also section \ref{vanet} for more insights about inter-vehicle communications) \cite{kanamori2012}.
Stigmergy is a mechanism of indirect cooperative communication among agents. In this case, vehicles are agents sharing their near-future plans as anticipatory stigmergies. Vehicles also re-schedule their routes based on the shared anticipatory stigmergies.
The study was based on simulations and six cases were modeled.
\begin{enumerate}
 \item {No-information} --- Vehicles find their best route using Dijkstra search before starting the trip, hence no traffic information is shared.
 \item {Long-term stigmergy} --- Similar to case $1$ (using Dijkstra search before departure), but based on long historical data (\textit{i.e.}, TT data) from probe vehicles equipped with GPS. Long--term data is updated daily.
\item {Long-term stigmergy} --- Similar to case $1$, but based on long historical data (\textit{i.e.}, TT data) from probe vehicles.
 \item {Combined Long-Term and short-term stigmergy} --- The latest five-minutes \textit{short} data is shared among vehicles along with the long-term data.
 \item {Anticipatory stigmergy} --- Vehicles share the links they intend to visit in the next ten minutes, and search for the best route based on a link performance function.
 \item {Anticipatory stigmergy with allocation strategy considering residual distance} --- An assignment method is used to assign drivers to the outputs from case $3$ and case $4$.
 \item {Anticipatory stigmergy with allocation strategy considering lost time of traffic congestion} --- Similar to case $5$, however based on the time spent in congestion.
\end{enumerate}
It is to be noted that methods from $1$ to $4$ can be seen as \textit{distributed} methods, where each vehicle locally decides the route to traverse. Nevertheless, the last two methods can be seen as \textit{centralized} methods, where a decision is taken by a decision-maker who assigns the vehicles to the routes.

\subsection{Discussion}

In Table \ref{sumOptim} we summarize the methods used to optimize urban traffic, namely traffic signal control and vehicle routing control, indicating the advantages and challenges of each method.
Traffic signal control optimizes the duration of signal-phases of each intersection to enhance traffic flow --- with fixed-time or adaptive time methods.
Due to flow variations and due to the need for real-time computations, the use of mathematical models to find a global optimal choice of green periods is a hard task. 
To overcome such a difficulty, distributed solutions are provided, while considering ''isolated'' intersections. As for vehicular routing control, the method aims at enhancing fleet management services, \textit{e.g.} pickup delivery services, as well as personal navigation solutions. The main challenge is to cope with the dynamism of the Vehicle Routing Problem (VRP), where new orders arrive arbitrarily, thus requiring a re-optimization process, in the case of fleet management. Regarding personal navigation systems, the major challenge is to find a fast algorithm with reduced memory cost.   

\begin{table*}[ht]
\centering
\scriptsize
\caption{Summary and comparison between the different approaches to UTO.}
\label{sumOptim}
\begin{tabulary}{0.001\textwidth}{ | l | c| c| c| }

  \hline    
  \hline      
  \textbf{Methods} & \textbf{Advantages} & \textbf{Disadvantages}  & \textbf{Articles}  \\ 
  \hline      \hline   
  
  Traffic signal  & It allows to regulate traffic flows, using a joint decision  & Flow variations and computation time limitations & \cite{Miller2009,Kouvelas2011,Fang2008}   \\
	control	        & on the duration of signal-phases associated to each     	   & make the provision of a global optimization &  \cite{Soares2012,Lin2011b,Lin2012}\\ 
  		           & signalized intersection in the road network.        	   & for the network a quite hard task. & \cite{deOliveira2010,Zegeye2012,Gokulan2010} \\

  \hline 
  Vehicle routing & Enhances fleet management operations, such as      & Fast algorithms are needed while using reduced    & \cite{Scellato2010}   \\
	control	   & courier pickup--delivery services or taxis.     	  & memory space. The dynamism of the VRP, due to   & \cite{Mohring2005,Pisinger2007}   \\
		   & Provide vehicle drivers with the most        	  & the new coming orders, renders the algorithms    & \cite{Pillac2012,kanamori2012}   \\
		   &  suitable ``optimal'' route.              	  & 	design more complex.	       &    \\			   
   \hline 
   \hline    
\end{tabulary} 
\end{table*}


\section{Vehicular Networks for UTM Applications} 
\label{vanet}

Inter-vehicle communications and ITSs are very related topics. ITSs greatly benefit from vehicle-to-infrastructure (V2I) and vehicle-to-vehicle (V2V) communications. 
When vehicles exploit V2V/V2I communications, they form a Vehicular Ad-Hoc Network (VANET) \cite{VANET}. Different from infrastructure-based networks, VANETs are formed on the fly. 
VANETs can be integrated with other types of networks, \textit{e.g.}, with WSNs (Wireless Sensor Networks) \cite{Barba2010}, or with cellular networks \cite{VanetCellular}. A seamless communication and integration between different types of networks makes the most benefit from all of them. Resource-constrained nodes of WSNs will be compensated by the vehicle nodes with no-energy consumption limitation (energy here is related to the energy needed for calculation processes). Data transmission challenges, \textit{e.g.}, packets loss rate, related to the highly mobile vehicle node can be compensated by the static WSN-nodes.  
The low penetration rate of communicating devices installed in the vehicles can be compensated via the large cellular coverage. 
In the coming subsections we will describe and review each of them, through illustrative examples from the literature.

\subsection{Architecture for VANET-based ITS}

The architecture of a VANET-based ITS network is principally composed of:

\begin{enumerate}
\item Road-Side Units (RSUs): transceivers fixed on the road side, to allow data exchange between the vehicles and the infrastructure;
\item On-Board Units (OBUs): in-vehicle transceivers, allowing the vehicles to communicate with each other (V2V) and with the infrastructure (V2I);
\item Traffic Management Center (TMC): a dedicated server where data are collected and decisions are taken;
\item Gateway: connects the backbone (TMC and RSUs) to the Internet, providing broadband connection to the users, and connects the VANET to a cellular network.
\end{enumerate}

\begin{figure}[ht] 
\centering
\includegraphics[scale=0.4]{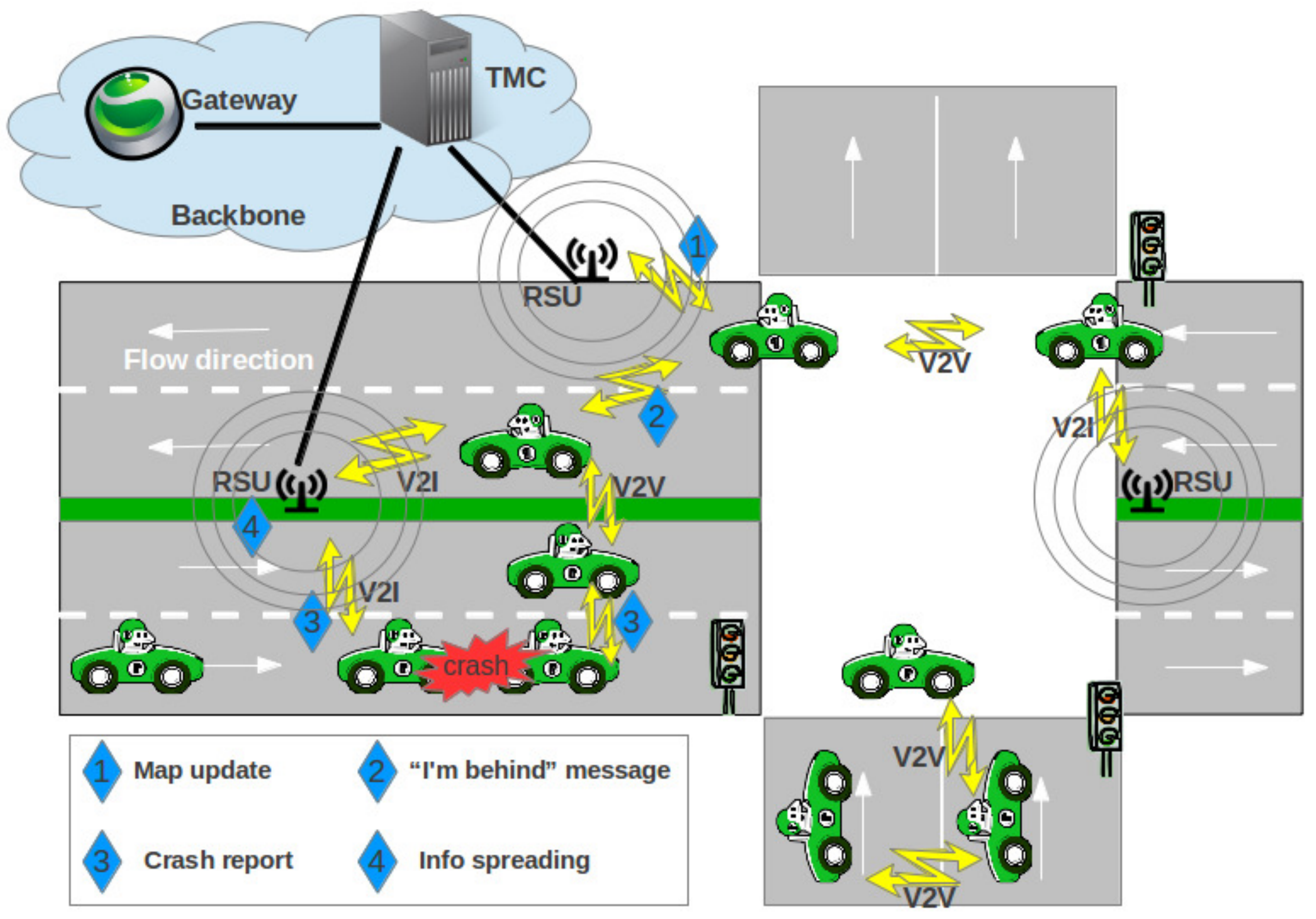}
\caption{Architecture of VANET-based ITS, composed of fixed RSU on the road-sides, OBU inside the vehicles, TMC and gateway for Internet connection.}
\label{fig:vanet1}
\end{figure}

We illustrate the architecture in Fig.~\ref{fig:vanet1}, with examples of exchanged messages over the V2I and V2V communications, for safety and navigation purposes. 
It is to be noted that the illustrated network is a \textit{hybrid} network, including both centralized and distributed aspects. 
The network is \textit{centralized} when the traffic information passes through the infrastructure to be processed at the TMC. 
However, the network can also be considered as \textit{distributed} when the data are collected, treated and re-distributed through V2V communications, without passing through the backbone \cite{Jerbi2007, Panichpapiboon2008, Garelli2011}.

DSRC (Dedicated Short Range Communication)/WAVE (Wireless Access in Vehicular Environment) is a major wireless-technology candidate for VANETs \cite{NotesDSRC10}.
DSRC/WAVE, standardized by IEEE 802.11p, is based on the WLAN IEEE 802.11a standard, where the ad-hoc mode is highly suitable to the vehicular mobile environment. Such environment is characterized by the frequent change that it causes to the network topology. 
DSRC/WAVE (IEEE 802.11p) has a similar PHY layer as IEEE 802.11a, with minor amendments, such as reduced channel-bandwidth to $10$MHz instead of $20$MHz. Nevertheless, major amendments to the MAC layer have been made to cope with the challenges of the mobile environment. 
The main objective is to increase data deliverability while decreasing the delay. Different from WLAN, a vehicle can send messages through a VANET without being associated to a particular service set. This property of VANETs has been achieved thanks to the replacement of the synchronization, authentication and association functions from the MAC layer to the higher layers.
A major MAC layer amendment with respect to WLAN is the \textit{multi-channel} service. A DSRC/WAVE device can listen to a channel while sending data on another channel using FDMA/TDMA.

In the USA, the spectrum band $5.850$ - $5.925$ GHz has been assigned to DSRC communications, while in Europe the band $5.725$ - $5.875$ GHz is assigned. In Japan $700$MHz band is assigned for next ITS system \cite{wave}. DSRC is meant to provide high data rate communication, as well as very low latency. DSRC is considered as a promising wireless standard which can be used to connect vehicles to infrastructure (V2I) and vehicle to vehicle (V2V). 
In table \ref{tab.dsrc} we report the main characteristics of IEEE 802.11p standard.
For further details about WAVE~/~IEEE 802.11p, the reader may refer to the tutorial article by Uzcategui and Acosta-Marum \cite{uzcategui2009wave}, and to the survey by Karagiannis \textit{et al.} \cite{karagiannis2011vehicular}.

\begin{table}[h!]
\caption{A summary of the characteristics of IEEE 802.11p standard}
\label{tab.dsrc}
\centering

\begin{tabular}{ | l | c| } 
  \hline      
  \hline
  \textbf{Characteristic} 	& \textbf{Values} 	 \\ 
  \hline      \hline   
  
  Range  	   	& up to $1000$ m 	\\
  Frequency band  	& $5.850$ - $5.925$ GHz \\
  Channelization  	& $1$ control channel and $6$ data channels  \\
  Channel bandwidth  	& $10$MHz 	   	\\
  Data rate  		& $6$ - $27$ Mbps 	\\
  MAC scheme 		& EDCA (Enhanced Distributed Channel Access)  	   \\
  Networking scheme 	& IPv6 \& WSMP(WAVE Short-Message Protocol)  	   \\

   \hline 
   \hline
\end{tabular} 
\end{table}

Research for VANETs implies the development of algorithms for routing, data dissemination, data aggregation, clustering, cluster head choice and mobility modeling.
Plenty of research papers proposed techniques, \textit{e.g.}, for data dissemination and routing, that ensure the delivery of information (packets) over the VANET \cite{Saleet2011}. 
Examples of data-dissemination techniques are broadcasting, flooding, request/reply, publish/subscribe, store-carry-forward.
Another point that has gained researchers' interest is the security in VANETs. It is to be noted that this paper is not targeting security issues. For more details, the reader is invited to consult the reference by Rivas \textit{et al.} \cite{Rivas2011}.

Although VANET applications are generally known to deal with safety-driving as well as drivers' comfort, we believe that less interest has been given to the concrete and specific application of urban traffic management.
In the following subsection, we provide a review about the existing work on VANET-based applications to UTM. The other subsections are devoted, respectively, to UTM solutions based on hybrid vehicular-sensor networks and vehicular-cellular networks.

\subsection{``Pure'' VANETs for UTM}
\label{vanet4utm}

A seminal VANET-based ITS was \textit{TrafficView} by Nadeem \textit{et al.} \cite{Nadeem2004}.
TrafficView was a hardware and software platform for finding the optimal route in a long trip or driving in situations like foggy weather. 
The TrafficView platform, implemented on a COMPAQ iPAQ PDA, allowed data sharing among vehicles on the road, to make the drivers aware of the road traffic. 
The design considers the minimization of transmitted data, in order to fit with the payload size, as defined in MAC IEEE 802.11b standard.
With the objective of providing real-time automatic route-scheduling for drivers, the authors presented and compared among different algorithms for vehicle records selection and aggregation.

Chen \textit{et al.} \cite{Chen2006} proposed a vehicular based networking and computing Grid (VGrid) framework that uses real-time position and velocity information of vehicles to smooth the vehicular flow. The information is exchanged through V2V communications. 
Similar to TrafficView, VGrid allows the drivers to ''see`` farther down the road and take an early action. 
For example, early changing of lane instead of waiting to reach a heavily congested lane because of a crash.
The proposed framework is fully distributed, where vehicles broadcasts messages containing their position and velocity on periodical basis. 
The first service offered by VGrid is an alert messaging which notifies about accidents, construction works or other obstruction in the road. 
Another service is the distributed computation of the Variable Speed Limit (VSL), a dynamic speed limit calculated based on traffic, and according to a linear function. 
An algorithm based on the local density of vehicles around each vehicle is used to determine the VSL. 
It has to be noted that, in this framework, the capacity of the grid computing is dynamic, increasing with the number of computing nodes, \textit{i.e.} the vehicles, thus increasing with the road congestion.
Different metrics were studied: 1) the speed variance normalized to the average speed, which is an indication of the amount of vehicles' acceleration and deceleration, 2) the throughput, defined as the number of vehicles exiting a road section of roadway in a fixed time interval, and 3) the latency, defined as the amount of time the vehicle takes to exit a road section. 
The objective is to minimize both variance and latency while maximizing throughput.

\begin{figure}[h]
\centering
\includegraphics[scale=0.25]{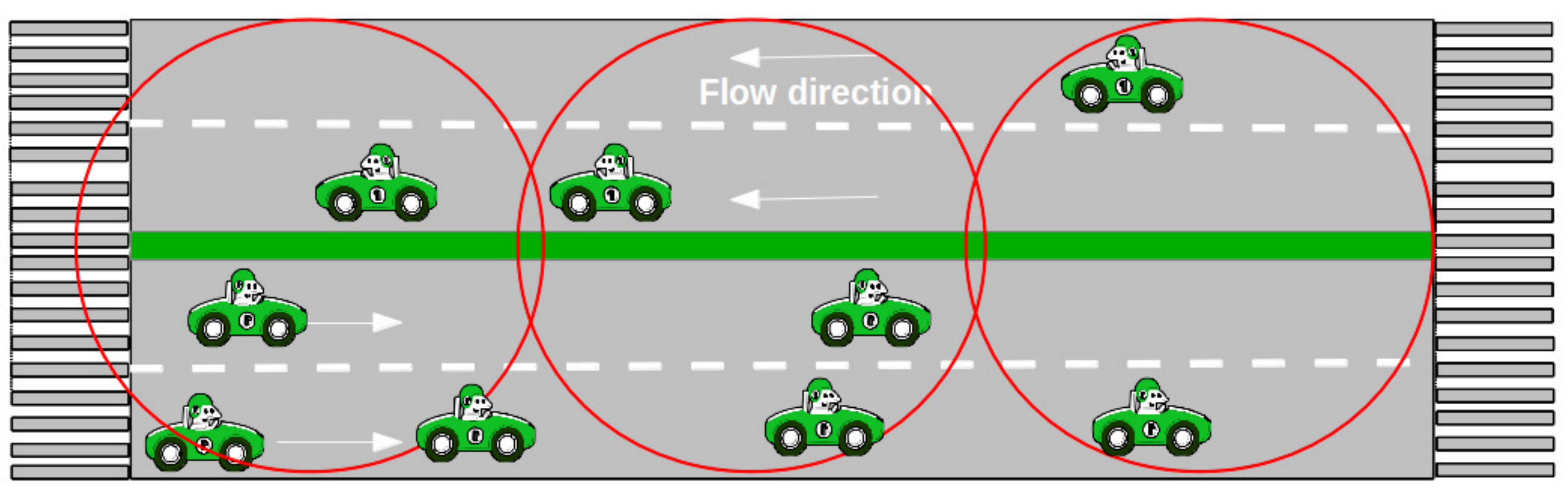}
\caption{A road segment of two directions divided into three cells according to the principle presented in \cite{Jerbi2007}.}
\label{fig:Jerbi}
\end{figure}

Jerbi \textit{et al.} \cite{Jerbi2007} proposed a cooperative decentralized algorithm to estimate traffic density in city roads (different from highways). 
Each road segment is divided into fixed-size cells (see Fig. \ref{fig:Jerbi}). The vehicles are supposed to know their own locations using a GPS receiver, as well as the coordinates of the cell centers.
While passing through the road segment, the nearest vehicle to the location of the cell center is elected \textit{leader} of the vehicles in the cell, for a given time period. 
Each vehicle maintains a table which contains position, velocity and direction of each neighbor vehicle. Such a table is updated periodically. Cell \textit{leaders} generate an estimation of the traffic, according to the information collected from the vehicles in their cells. 

Another distributed approach strictly benefiting from V2V communications was proposed by Panichpapiboon and Pattara-Atikom \cite{Panichpapiboon2008}. The authors considered a probe vehicle that chooses the best estimate for the vehicle density, based on the number of neighbors in its vicinity. The authors found an expression showing that if the inter-vehicle spacing is exponentially distributed, then the local density is the best estimate for the global vehicle density. Unlike the solution proposed by Jerbi \textit{et al.} \cite{Jerbi2007}, this one does not depend on a fixed cell size, but considers the maximum transmission range of the OBU.
The authors analyze two schemes: one-hop neighbor scheme and two-hops neighbor scheme. In the first scheme, the estimation of the vehicular density 
is obtained using a Poisson process, where the estimated vehicular density is a function of the transmission range of the installed OBU. In the two-hops neighbor scheme, Poisson process is also used, with the range of the one-hop case replaced by the extended range caused by the vehicle acting as a relay.

Garelli \textit{et al.} proposed a distributed V2V-based method to estimate urban traffic density in real-time, for a target area \cite{Garelli2011}. In the proposed method, the sampler vehicle plays two roles: collecting data for urban density estimation and handing over the task to another vehicle, when about to leave the target area.
The authors do not specify how the sampler is chosen, however they emphasize on the importance of the sampler choice and on the impact it causes on the density estimation. Randomization of the distance between the old and the new sampler is adopted. 
Differently from the references presented above, \cite{Jerbi2007, Panichpapiboon2008}, the algorithm does not depend on a fixed-size cell nor adopts one/two-hop schemes. Instead, it uses the geocasting method introduced by Borsetti \textit{et al.} \cite{geocasting}. 
At uniformly distributed random sampling instants, the sampler broadcasts a POLL message. Once the message has been received by the vehicles located in the region of interest, the vehicles respond back to the sampler with a REPLY message. The sampler then uses the number of REPLY messages to estimate the vehicle density. 

Ma \textit{et al.} \cite{Ma2009} propose an algorithm based on Support Vector Machine (SVM) to characterize road traffic (\textit{i.e.} highway) conditions and to detect incidents. The algorithm is designed to be executed by the vehicles and the output is to be sent to the infrastructure on periodic bases, every $s_t$ time step. 
The algorithm's output is transmitted (along with the time stamp and the vehicle location), either via a direct connection with a ''near`` RSU, \textit{i.e.} V2I, or via relying the output information over V2V connections till it reaches an RSU. 
Each vehicle classifies its traveling experience, using SVM, using three categories:
\begin{enumerate}
 \item normal, no-accident scenario, coded as $\{-1\}$,
 \item passed by accident location, coded as $\{+1\}$,
 \item stopped in a queue, coded as $\{+2\}$.
\end{enumerate}
This classification is based on the speed variation and the lane changing behavior of the vehicle.

The RSU determines the final decision regarding the status of the highway traffic, based on the outputs of the SVM algorithm, which are sent by vehicles equipped with OBUs.
Each RSU supervises a section of the highway, where each section is divided into segments. The number of segments $n_i$ within
a section $i$ depends upon the length of the section $L_i$, the average travel speed $\bar{v_i}$, and the period $s_t$ among two 
consecutive message-sending. The number of segments was then determined as $n_i = L_i/(\bar{v_i}.s_t)$.
The authors evaluated the proposed method using microscopic simulations. 
Based on the highway network of Spartanburg, SC--USA, incidents are simulated by blocking one, two or three lanes while recording the data generated by the vehicles, \textit{i.e.} speed variations and lane changing status.
The SVM algorithm was compared to a California \#7 incident-detection algorithm \cite{cali7}. 
The detection time, the detection rate and the false-alarm rate are the metrics used to compare between California \#7 and SVM algorithms. 
The performance of SVM exceeds the California \#7 algorithm in terms of detection and false-alarm rates, while both algorithms achieve $100$\% accident detection rate.

The estimation of urban traffic density is an important factor also with respect to data routing in vehicular networks --- routing path decisions are considerably affected by such a density. Although the focus of our survey is not on data routing protocols in VANETs, we emphasize the significance of road traffic conditions, and hence of the potential presence of relaying nodes, on designing efficient routing protocols in VANETs \cite{Rondinone2010}. 

Landmark Overlays for Urban Vehicular Routing Environments (LOUVRE) \cite{Lee2008} and Road-Based using Vehicular Traffic routing (RBVT) \cite{Nzouonta2009} are proactive geo-routing protocols that use real-time road traffic estimations in order to choose the forwarding routes that enhance end-to-end data deliverability. In these algorithms, the vehicles exchange periodical messages with the target of drawing a connectivity map. The vehicles use such a map to take a decision on the most convenient path that ensures data delivery to the destination.
A more recent contribution was proposed by Saleet \textit{et al.} \cite{Saleet2011}, for an intersection-based geographical routing protocol that guarantees network connectivity among road intersections, taking into account QoS constraints on tolerable delay, bandwidth usage and error rate. For more details about data routing in VANETs, the reader can refer to the survey reference by Chen \textit{et al.} \cite{Chen2009}.

\subsection{Hybrid Vehicular-Sensor Networks}

Wireless Sensor Networks (WSNs) are widely used mainly due to their cost effective advantage. On the other hand, WSNs are limited with respect to resources, \textit{i.e.} energy, bandwidth and storage space.
WSNs are mainly used in industrial applications and in safety/healthcare systems. ITS applications have also gained interest for WSN \cite{wsn-4-its} \cite{wsn-4-its2}. For example Herrera-Quintero \textit{et al.} \cite{wsn-4-its2} presented a case study where WSNs are used to implement a parking management system. 

Vehicular networks can be considered as Mobile WSNs (MWSNs), with the advantage of non-applicable energy constraint, unlike the WSN energy-scarce case. Another advantage of VANETs with respect to MWSNs is the localization issue. 
Most of the works that address localization in WSNs/MWSNs present algorithms based on received signal strength. 
However, VANETs may use in-vehicle GPS receivers, as well as V2V cooperative communications \cite{yao2011improving}, to further improve localization accuracy.

Mainly motivated by the the cost-effective advantage of sensors and in order to overcome the challenge of low data deliverability in mobile environment, hybrid vehicular-sensor networks have been proposed \cite{Barba2010} \cite{HuaQin2010} \cite{vanetWSN2}. 
The idea is to mix both networks to extend the coverage of VANETs and to ensure higher data deliverability, especially when the vehicle density is low. 
Nevertheless, integrating both networks raises heterogeneity contests: VANETs nodes have higher processing power while WSN nodes are throughput-limited and less mobile compared to VANET nodes. 

An interesting framework of VANET-WSN integration was presented by Qin \textit{et al.} in \cite{HuaQin2010}, with focus on rural highways.
The concern of the authors is about providing safety-driving through data exchange between the two networks. Motivated by the cost-effective WSN, sensor nodes are to be placed on the roadside
giving data access to a solo passing vehicle, \textit{i.e.}, a vehicle not associated with any cluster (of other neighbor vehicles).
Sensor nodes form neighboring groups which communicate with each other through a common gateway sensor node, acting as an Access Point (AP).
In the proposed system, vehicles are supposed to be equipped with two communication interfaces: one interface to communicate with other vehicles, 
and a second interface (e.g. ZigBee IEEE 802.15.4 interface) to communicate with sensor nodes on the roadside. 
Each AP broadcasts a beacon message with safety-related information on periodical bases. Passing vehicles (\textit{i.e.} cluster heads) send registration requests to the AP, upon hearing the beacon messages.
Due to energy limitation in sensor nodes, packets are transmitted only when events (\textit{e.g.}, a deer roaming on the road) are detected or when a vehicle (\textit{i.e.}, cluster head) enters the coverage zone of the AP.
In addition, the authors proposed a TDMA-based protocol to support real-time data delivery while saving energy. Two main challenges are addressed: 
\textit{intra-group scheduling}, whose objective is to assign the sending and receiving slots, while reducing the packet loss rate within the sensor node; 
\textit{inter-group scheduling}, which addresses how the AP successfully delivers packets from one group to another with low delay. 

To extend the geographical scope of VANETs, with the aim of improving road safety as well as informing drivers with traffic density and weather conditions, was the concern of Barba \textit{et al.} \cite{Barba2010}. In this sense, the authors presented a simple communication protocol between a source sensor node and a sink vehicle.
The sink vehicle belongs to a cluster of vehicles that is passing through the coverage area of the sensor node.
The authors evaluated the performance of the hybrid network (vehicular and sensor nodes) through simulations. Based on the data amount needed to communicate the status of a specific number of road segments, the coverage range and the average speed of the vehicle, the connectivity time (between the sensor node and the passing vehicle) is calculated. The performance in terms of throughout, end-to-end latency and packet loss is evaluated under two well-known routing protocols for VANETs: Ad hoc On Demand Distance Vector (AODV) and Dynamic Source Routing (DSR).

 \subsection{Hybrid Vehicular-Cellular Networks}

What makes communications among vehicles challenging is the dynamic topology of VANETs. Consequently, developing data routing algorithms is complex, especially when the target destination is not near the source and multi-hop communication is needed. The low penetration rate of OBUs has a negative impact on the system efficiency. A reduced number of participating nodes in data sharing/transmission makes delivery to the destination more unreliable. 

Cellular networks have been proposed to provide backhaul architecture facing the low penetration rate or low vehicle density situations. Hence data, in particular non-safety related data, which are non-critical, can reach their destination. 

Santa \textit{et al.} emphasized the potential of cellular networks to overcome the complexity of routing protocols in a dynamic-topology environment \cite{vanet.cell01} \cite{vanet.cell02}.
The authors proposed a high level communication architecture, where vehicle groups are formed. Vehicles exchange messages using P2P communication over a 3G cellular network.
Traffic zones are formed and divided in coverage areas, where each area uses a different P2P communication group. The coverage areas are related to the proposed architecture and not to be mixed with the cellular coverage areas.
An Environment Server (which is equivalent to a RSU), manages the messages exchange inside the area. Two transmission scenarios are possible: either P2P messages are sent to a specific vehicle, or P2P messages are broadcasted within the area.
All related information about each area, \textit{e.g.}, its available services, are maintained in an entity called Group Server. The environment server receives and sends information from/to the group servers (in a hierarchical manner).

Lin \textit{et al.} proposed a strategy for merging V2V data with Floating Car Data (FCD) \cite{Lin2006}, with the objective to enhance the precision of road traffic maps by increasing the number of sampling cars, while overcoming the low penetration rate of both OBUs and cars with FCD collectors, the privacy issue and the FCD data filtering processes.
Vehicles are grouped into three categories: FCD vehicles, equipped with FCD tools, V2V-vehicles, equipped with OBUs, and master vehicles, equipped with both systems. A clustered sub-network is formed between a master vehicle and the neighbor V2V vehicles, where the former plays the role of the cluster head. The master vehicle collects floating data (\textit{e.g.} position, speed) from V2V vehicles and sends it periodically to the TMC.
Communication between the V2V vehicles and the master vehicle is performed by means of an unicast routing protocol.

In the context of mobile communications for vehicular environment, LTE-Advanced is probably the most promising wireless broadband technology enabling a wide range of applications, while providing enhanced network performance, \textit{e.g.}, capacity, spectral efficiency and interference management \cite{Ali2013}. Peak target data rates are expected to be $1$ Gb/s in the downlink and $500$ Mb/s in the uplink. LTE-A, Rel. $12$, is supporting D2D (Device to Device) communication, where devices are able to communicate in ad-hoc mode without the need to connect to the base station. It is worth mentioning that D2D communication over LTE air-interface (3GPP) is becoming a strong potential competitor to IEEE $802.11$p, for vehicular communication.

As we talk about integration of VANET with cellular networks, we should mention the activity of the ISO TC204 WG16, on the standardization of the Continuous Air Interface for Long and Medium range (CALM). The standard aims at integrating additional interface protocols --- GSM/GPRS (2/2.5G), UMTS (3G), infrared communication and wireless systems in $60$ GHz band --- on top of the IEEE 802.11p \cite{calm01,calm03}.
Within the Cooperative Vehicle-Infrastructure Systems (CVIS) project, a CALM platform has been developed and tested \cite{cvis}, where IPv6 protocol is used to perform various types of P2P communications on different communication channels.

\subsection{Discussion}
Table \ref{tab:compareVANET} summarizes the pros and cons of the three illustrated vehicular network approaches to UTM. Clearly, a pure VANET approach, based on a homogeneous network, has the advantage of dealing with a less complicated data management process, with respect to hybrid vehicular networks. However, in conditions of low vehicle density, data deliverability is hard to be guaranteed. 
Moreover, data routing in highly dynamic environments is not a straightforward task. Integrating VANETs with other networks, \textit{i.e.}, WSN or cellular (in particular, LTE-based), helps to overcome the coverage limit, when the traffic density is low. Furthermore, data exchange over VANETs is cost-effective with no fees, unlike the case of cellular networks, where data dissemination depends on the user data plan.

In the heterogeneous network scenarios, such as the case of hybrid vehicular-sensor network, the main challenge is data management. For example, coping with the mismatch between the high processing rate of vehicle nodes and the reduced capabilities (\textit{e.g.}, in terms of bandwidth and energy) of the light sensors.  
Although the roadside sensor nodes, in hybrid vehicular-sensor, can be seen as equivalent to the RSUs of a homogeneous VANET, the motivation for such a replacement is mainly due to the fact that sensor nodes are lighter, hence more cost-effective, than RSUs.

Hybrid vehicular--cellular networks, provides a good solution to the coverage issue, being able to reach a high number of devices also in case of low density scenarios. Open issues are the data exchange cost over the cellular infrastructure, and the latency of data delivery, which is higher than the two previously discussed solutions.
Consequently, this third option is not suitable for safety-driving applications, but could be efficiently applied in many useful application scenarios, where latency does not represent an issue, and the distance from the event or a generic geo-localized information is sufficient to guarantee a proper data dissemination.
 
\begin{table*}[ht]
\centering
\scriptsize
\caption{Summary and comparison among the different approaches of vehicular networks for UTM applications}
\label{tab:compareVANET}
\begin{tabular}{ | l | c| c| c| } 
  \hline    
  \hline      
  \textbf{Methods} & \textbf{Advantages} & \textbf{Disadvantages}  & \textbf{Articles}  \\ 
  \hline      \hline   
  ``Pure'' VANETs  & Homogeneous networks based on    & Low coverage in case of low traffic density.   & \cite{Nadeem2004, Chen2006} \\
			             & DSRC/WAVE technology, that provide 		& Data delivery is hard to guarantee in highly 		&\cite{Jerbi2007, Panichpapiboon2008}	\\
			             & low-latency end-to-end data delivery.	&  dynamic environment.			  & \cite{Garelli2011, Ma2009}  \\
  \hline 
  Hybrid Vehicular Sensor  & Sensor nodes are light, cheap and & Heterogeneity between vehicular and      & \cite{Barba2010}  \\
   	Networks      	      & have low energy consumption.	 & sensor networks. Resource limitations:         & \cite{wsn-4-its, wsn-4-its2} \\ 
					                & No charges for data transmission. &  \textit{i.e.} energy, bandwidth and storage. &	\cite{vanetWSN2, HuaQin2010}\\
   \hline 
  Hybrid Vehicular Cellular & They guarantee a 	   	& High latency and higher energy consumption than   & \cite{vanet.cell01,vanet.cell02} \\ 
	Networks	               & 	very good coverage.   & the two previous options. 			                  & \cite{Lin2006,cvis} \\
			                     & 	            	    		& Not suitable for safety-related applications. 	  & \\
   \hline
   \hline    
\end{tabular} 
\end{table*} 


\section{Future research directions and challenges}
\label{model}

Achieving efficient and cost-effective UTM requires reliable data collection/estimation, proper modeling and efficient data dissemination.
Despite the fact that there is a huge amount of high-quality research works tackling UTM (\textit{i.e.}, traffic estimation and optimization), two main challenges remain:
\begin{itemize}
 \item dealing with large networks using time-efficient algorithms,
 \item collecting fine-grain data while preserving the privacy of the drivers.
\end{itemize}

Methods based on Markov modeling or Bayesian networks, defined for a particular road network, provide near-optimal solutions. 
Nevertheless, they require high performance computing, especially if the proposed model considers a large network, causing the increment of the number of states.  
Other methods based on heuristics, such as Tabu search or cellular automata, are more efficient in terms of execution time, at the cost of reduced accuracy and output optimality.

Concerning the privacy issue, distributed solutions such as pure VANET approaches and P2P Internet-based overlays (\textit{e.g.} Distributed Geographic Table  \cite{Picone2011a}) have the native advantage of evenly distributing the load of the anonymously collected information and the responsibility to maintain it.
Hence avoiding the risk to centralize all the harvested data in one single knowledge repository.

With the progress in communication technologies, cellular systems, satellite communications and in particular VANETs, data collection and dissemination is becoming more comfortable.
VANET functionalities, \textit{e.g.}, clustering and data routing, have been remarkably addressed by researchers. 
Nevertheless, effective (and widely deployed) UTM applications based on vehicular networks still have to come.
It is also to be noted that most of the few works tackling VANET-based UTM \cite{Chen2006} \cite{Jerbi2007} \cite{Garelli2011} used of heuristic approaches.

Based on the previous inputs, synthesized below:
\begin{itemize}
\item the complexity of solving analytical models for large networks,
\item privacy-related issues, concerning anonymous data collection and storage in a single knowledge repository,
\item the fast evolution of inter-vehicle communications, hence the potential of using VANET architectures with highly distributed algorithms, 
\end{itemize}
there is a need, and potential, to further investigate the exploitation of the VANET framework, in order to propose distributed algorithms for the time-consuming models (either Markovian or Bayesian).
Such distributed techniques are expected to provide near-optimal global solutions. To this purpose, the following steps may be included:
\begin{enumerate}
\item developing a "simple" Markovian model, for one intersection, \textit{e.g.} the approach proposed by Osorio \textit{et al.} \cite{Osorio2009},
\item generalizing the model to be able to deal with a large network, while identifying the most time-consuming calculation tasks,
\item making use of the VANET architecture, as well as distributed techniques (\textit{e.g.} distributed machine learning or Game Theory) to overcome the time consuming calculations --- for example, by placing time consuming tasks at the RSU.
\end{enumerate}
As examples of novel UTM applications, we foresee the usage of VANETs to solve the best-route problem for drivers, whether for fleet management services or for individual drivers. Best-route algorithms can also be adapted to solve crisis/evacuation situations.

On the other hand, hybrid vehicular networks is quite a recent topic. Few works exist and there is still some way to go in this area. 
In the context of hybrid vehicular--sensor networks, an open research direction is the development of distributed methods for data aggregation/dissemination, while:
\begin{itemize}
\item considering two types of data transmissions: data transmission from a static sensor node to a mobile vehicle node and vice versa;
\item assuming the vehicles have full knowledge about all locations of static sensor nodes (e.g., in a region), based on the fact that OBUs have the storage capability;
\item placing the tasks demanding higher processing power at the vehicle nodes.                                                                                         
\end{itemize}
Concerning hybrid vehicular-cellular networks, the main motivation behind the idea of integrating VANETs and cellular networks is to provide backhaul coverage, hence overcoming the low vehicle density problem.
Nevertheless, the high cost of data transmission over the cellular network is still an open issue. New models are needed to incentivize the cellular operators for the integration of VANETs with their networks.
In this direction, the following research points are still open: 
\begin{itemize}
\item 
new business models that consider the pricing plans of cellular users, with the target of incentivizing the cellular operators to participate in ITS applications;
\item
inter-RAT (Radio Access Technology) selection algorithms for the transmitting OBUs, to allow for a proper choice of their most suitable receiving network, whether the VANET (\textit{i.e.} another OBU or RSU) or the cellular network. 
\end{itemize}
It is worth mentioning that the inter-RAT selection algorithms should consider; the cost of data transmission, the probability of delivering packets to the destination and the end-to-end delay. 

 
\section{Conclusion}
\label{conc}

In this survey, we reviewed and classified recent research works on the topic of urban traffic management. 
Estimation/collection of urban traffic information is performed by means of five approaches: analytical methods, fixed sensors, mobile phones, mobile phones combined with GPS receivers, and hybrid methods.
We described such approaches and provided examples from the literature for each of them, highlighting their advantages and disadvantages.
As for the urban traffic optimization techniques, we grouped them into two major categories: traffic signal control and vehicle routing control,
which we illustrated referring to the literature.  
Afterwards, we emphasized the role of communication technologies, in particular those enabling vehicular networks, which have been used by the research community to propose highly decentralized intelligent transportation systems.
After a brief introduction to VANETs, we divided the architectures serving decentralized ITS applications into three groups: pure-VANETs, hybrid vehicular-sensor networks and hybrid vehicular-cellular networks. We clarified the motivation of the hybrid architectures, and we discussed their pros and cons. 
Finally, we concluded our survey by illustrating the main research challenges of the reviewed methods, also providing future research directions.


\bibliographystyle{elsarticle-num}

\end{document}